\documentclass[12pt]{article}
\usepackage[utf8]{inputenc}
\usepackage[top=50pt,bottom=50pt,left=68pt,right=66pt]{geometry}
\usepackage{amsmath}
\usepackage{booktabs}
\usepackage{amssymb}
\usepackage[title]{appendix}
\usepackage{mathrsfs}
\usepackage{cite}
\usepackage{float}
\usepackage{xcolor}
\usepackage{multirow}
\usepackage{graphicx,caption,subcaption}
\usepackage[space]{grffile}
\newcommand{\overbar}[1]{\mkern1.5mu\overline{\mkern-1.5mu#1\mkern-1.5mu}\mkern 1.5mu}

\definecolor{darkblue}{rgb}{0.1,0.1,.7}
\usepackage[breaklinks=true,linktocpage=true,
colorlinks=true,urlcolor=darkblue,linkcolor=darkblue,
citecolor=darkblue,pdfpagelabels=true,hypertexnames=true,
plainpages=false,naturalnames=false,]{hyperref}

\usepackage{datetime}
\newdateformat{monthyeardate}{%
  \monthname[\THEMONTH], \THEYEAR}
\date{\monthyeardate\today}

\numberwithin{equation}{section}

\def\ca{{\cal A}}
\def\cb{{\cal B}}
\def\cc{{\cal C}}
\def\cd{{\cal D}}
\def\ce{{\cal E}}
\def\cf{{\cal F}}

\def\ch{{\cal H}}

\def\cl{{\cal L}}

\def\cn{{\cal N}}
\def\co{{\cal O}}

\def\car{{\cal R}}

\def\cw{{\cal W}}

\def\bq{{\bf Q}}

\def\bs{{\bf S}}
\def\bt{{\bf T}}

\def\bv{{\bf V}}

\def\bx{{\bf X}}

\newcommand{\al}{{\alpha}}
\newcommand{\dal}{{\dot{\alpha}}}

\begin{document}

\renewcommand{\arraystretch}{1.3}
\thispagestyle{empty}

{\hbox to\hsize{\vbox{\noindent\monthyeardate\today}}}

\noindent
\vskip2.0cm
\begin{center}

{\Large\bf Aspects of cubic nilpotent superfields}

\vglue.3in

\hspace{-2pt}Yermek Aldabergenov,${}^{a,b,}$\footnote{yermek.a@chula.ac.th} Ignatios Antoniadis,${}^{c,d,}$\footnote{antoniad@lpthe.jussieu.fr} Auttakit Chatrabhuti,${}^{a,}$\footnote{auttakit.c@chula.ac.th} Hiroshi Isono${}^{a,}$\footnote{hiroshi.isono81@gmail.com}
\vglue.1in

${}^a$~{\it Department of Physics, Faculty of Science, Chulalongkorn University,\\ Phayathai Road, Pathumwan, Bangkok 10330, Thailand}\\
${}^b$~{\it Department of Theoretical and Nuclear Physics, 
Al-Farabi Kazakh National University,\\ 71 Al-Farabi Ave., Almaty 050040, Kazakhstan}\\
${}^c$~{\it School of Natural Sciences, Institute for Advanced Study, Princeton NJ 08540, USA}\\
${}^d$~{\it Laboratoire de Physique Th\'eorique et Hautes Energies (LPTHE), Sorbonne Universit\'e,\\ CNRS, 4 Place Jussieu, 75005 Paris, France}\\
\vglue.1in

\end{center}

\vglue.3in

\begin{center}
{\Large\bf Abstract}
\vglue.2in
\end{center}

We study in more detail the cubic constraints for $\cn=1$ chiral superfields proposed in the earlier work {\it Eur. Phys. J.} C 81, 523 (2021), which describe low-energy goldstino-axion dynamics in global non-linearly realized supersymmetry. We generalize the constraint (i) by gauging the abelian symmetry that shifts the axion, and (ii) to supergravity. In the former case we show that after adding the abelian gauge multiplet, the construction can be formulated as a cubic nilpotent massive vector superfield. In the latter case we find explicit solution to the constraint including contributions from the supergravity multiplet. We also propose microscopic supergravity models (with linear supersymmetry), including a toy model for the string dilaton, which give rise to the cubic constraints at low energies.

\newpage

\tableofcontents

\setcounter{footnote}{0}

\section{Introduction}

Constrained superfield approach to non-linear supersymmetry (SUSY) is a powerful tool which can be used to obtain effective supersymmetric field theories operating at energies below the SUSY breaking scale \cite{Rocek:1978nb,Ivanov:1978mx,Lindstrom:1979kq,Ivanov:1982bpa,Samuel:1982uh,Casalbuoni:1988xh,Komargodski:2009rz}. The simplest and well-known example uses a quadratic nilpotent chiral superfield $\bs^2=0$, which removes the complex scalar component $S$ as $S=\chi^2/(2F)$,~\footnote{We use spinor notations of Wess and Bagger \cite{Wess:1992cp}, in particular $\chi^2\equiv\chi^\alpha\chi_\alpha$.} where $\chi$ is its fermionic partner (playing the role of the goldstino), and $F$ is the complex auxiliary scalar which must be non-zero, since otherwise the solution is singular. After applying the constraint, the resulting action describes a goldstino of non-linearly realized $\cn=1$ SUSY, and can be related \cite{Kuzenko:2010ef} to the original Volkov--Akulov theory \cite{Volkov:1973ix}. If the goldstino is accompanied by other (light) fields, these fields can be described by introducing additional superfields and imposing appropriate orthogonality constraints \cite{Komargodski:2009rz,Ferrara:2015tyn,Kahn:2015mla,DallAgata:2016syy,Cribiori:2017ngp}. Constrained superfields were used in particular in supergravity-based cosmology, for example in constructing de Sitter vacua \cite{Antoniadis:2014oya,Bergshoeff:2015tra,Hasegawa:2015bza,Kuzenko:2015yxa,Kallosh:2015tea,Ferrara:2015gta,Schillo:2015ssx,Bandos:2016xyu,Farakos:2016hly,Cribiori:2016qif,Dudas:2015eha,Antoniadis:2015ala,DallAgata:2015pdd} as well as inflationary model building \cite{Antoniadis:2014oya,Ferrara:2014kva,Kallosh:2014via,Kallosh:2014hxa,DallAgata:2014qsj,Kahn:2015mla,Ferrara:2015tyn,Carrasco:2015iij,DallAgata:2015zxp,Dudas:2016eej,Delacretaz:2016nhw,Argurio:2017joe}.

In the preceding work \cite{Aldabergenov:2021obf} we studied a new type of cubic constraints for $\cn=1$ chiral superfields (in global SUSY), and showed that its solutions can describe a goldstino interacting with an axion, with non-linearly realized supersymmetry (see also \cite{Aldabergenov:2021rxz} for an $\cn=2$ generalization of this constraint). This cubic constraint can be written in the form $(\bs+\overbar\bs)^3=0$, and its leading component $S+\overbar S$ is eliminated in terms of a real (at least) bilinear function of the goldstino $\chi$ and $\bar\chi$. The axion in this case is the imaginary part of $S$, and is protected by its global shift symmetry (spontaneously broken everywhere in field space). Alternatively, if an abelian symmetry is realized linearly, as a phase rotation of $S$, the cubic constraint can be written as $(\bs\overbar\bs-\upsilon^2)^3=0$ (assuming that $\upsilon\equiv\langle S\rangle\neq 0$), in which case the axion is the angular part of $S$, while the radial part is eliminated by the constraint. As shown in \cite{Aldabergenov:2021obf} these two cubic constraints (and their solutions) can be related by a simple field redefinition $S\rightarrow\log S$ (from shift-symmetric case to phase-symmetric), and therefore in this paper we will focus on the shift-symmetric case for simplicity.

In Ref. \cite{Aldabergenov:2021obf} we also showed that the solution to the cubic constraint is related by non-linear field redefinitions to the orthogonal nilpotent superfields proposed in \cite{Komargodski:2009rz} that preserve the same degrees of freedom (goldstino and axion). The orthogonality constraint can be written as $\bx(\bt+\overbar\bt)=0$, where $\bx$ is the quadratic nilpotent goldstino superfield, while $\bt$ is the orthogonal superfield whose only surviving component is the imaginary scalar (axion). The solution to our cubic constraint can be related to the solution to the orthogonality constraint by using the fact that the combination $\bq\equiv\log(\bx+e^\bt)$ contains the same independent degrees of freedom as $\bs$ (which is constrained as $(\bs+\overbar\bs)^3=0$), and is cubic nilpotent, $(\bq+\overbar\bq)^3=0$. For this relation to work, we also require that under the abelian symmetry $\bq$ transforms by an imaginary shift, i.e. $\bx$ and $e^\bt$ transform identically, by a phase rotation. In other words, a Lagrangian of $\bs$ is equivalent to a particular Lagrangian of $\bx$ and $\bt$ (respective constraints assumed), where they form the combination $\bx+e^\bt$.

One possible advantage of using the cubic constraint as opposed to the orthogonal nilpotent supefields was discussed in Ref. \cite{Terada:2021rtp} in the context of minimal supergravity inflation (MSI), where only inflaton and goldstino fields are present \cite{Ferrara:2015tyn,Carrasco:2015iij}. It was shown in \cite{Hasegawa:2017hgd} that MSI with the orthogonal nilpotent superfields generally suffers from catastrophic (and possibly pathological) production of slow gravitinos due to changing sound speed (unrelated to the standard gravitino problem), but Terada \cite{Terada:2021rtp} showed that if we instead construct MSI with a single chiral superfield subject to the cubic constraint, the gravitino sound speed does not change, and the problem is avoided. Cosmological applications of the cubic constraints are further discussed in \cite{Aoki:2021nna}.

In this work our goal is to derive explicit generalizations of the cubic constraint to the cases of (i) local abelian symmetry by gauging the axion shift, and (ii) local supersymmetry.~\footnote{When local SUSY is spontaneously broken, we can choose the super-unitary gauge where the goldstino vanishes, and the solution to the constraint $(\bs+\overbar\bs)^3=0$ is simply $S+\overbar S=0$ (if goldstino is purely the fermionic component of $\bs$), which can dramatically simplify phenomenological applications of these constraints. In this work we will derive explicit solutions prior to the gauge fixing.} In the former case we introduce a real gauge superfield which in the unitary gauge absorbs the chiral (St\"uckelberg) superfield $\bs$ and becomes massive. The cubic constraint can then be written as a cubic nilpotency condition for the massive vector superfield. We also construct microscopic supergravity (SUGRA) models with linear SUSY, which lead to an appropriate cubic superfield constraint upon integrating out a heavy scalar (saxion).

This paper is organized as follows. In Section \ref{sec_cubic_rigid} we review the cubic constraint for chiral superfield with global shift symmetry, and generalize the results to the case of gauged shift symmetry and obtain cubic nilpotent massive vector superfield. In Section \ref{sec_cubic_SUGRA} we generalize the results of Section \ref{sec_cubic_rigid} to supergravity, and for each constraint we construct a minimal Lagrangian. In Section \ref{sec_UV} we discuss several examples of microscopic models for the cubic constraints, including R-symmetric models and toy models for the string dilaton.

\section{Cubic constraints in rigid SUSY}\label{sec_cubic_rigid}
\subsection{Constrained chiral superfield}

Consider an $\cn=1$ chiral superfield $\bs$,
\begin{equation}
	\bs=S(y)+\sqrt{2}\theta\chi(y)+\theta^2F(y)~,
\end{equation}
where $S$ is a complex scalar, $\chi$ is a Weyl fermion, and $F$ is auxiliary complex scalar (the expansion is in chiral basis, $y^m=x^m+i\theta\sigma^m\bar\theta$).

Assuming the presence of a shift symmetry, $S\rightarrow S+ic$ ($c$ = real constant), let us impose the following invariant cubic constraint,
\begin{equation}
	(\bs+\overbar\bs)^3=0~.\label{cubic_constr}
\end{equation}
We parametrize $S$ as
\begin{equation}\label{S_param}
	S=\tfrac{1}{2}(\phi+i\varphi)~.
\end{equation}
and assume the VEV $\langle\phi\rangle$ is zero. $\varphi$ is the axion, shifting by a constant under the abelian symmetry. 

The highest component of \eqref{cubic_constr} reads
\begin{equation}\label{cubic_constr_comp}
    \tfrac{1}{4}\phi^2\Box\phi+\phi B=H~,
\end{equation}
where $B$ and $H$ are defined as follows,
\begin{align}\label{BH_def}
\begin{aligned}
B &\equiv
	2F\overbar F-\tfrac{1}{2}\partial\varphi\partial\varphi-\left(i\chi\sigma^m\partial_m\bar\chi+{\rm h.c.}\right)~,\\
	H &\equiv
	\chi^2\overbar F+\bar\chi^2F+\chi\sigma^m\bar\chi\partial_m\varphi~.
\end{aligned}
\end{align}
Eq.~\eqref{cubic_constr} can be solved by utilizing the property $H^3=0$ and $\phi^3=0$ (the former is a consequence of the fermionic property $\chi^3=0$, and the latter is the leading component of \eqref{cubic_constr}), and the solution reads \cite{Aldabergenov:2021obf}
\begin{equation}\label{phi_sol}
    \phi=\frac{H}{B}-\frac{H^2}{4B^4}\Box H~,
\end{equation}
At the leading order in $\chi$ and $\bar\chi$ the solution reads
\begin{equation}\label{phi_sol_leading}
	\phi=\frac{\chi^2\overbar F+\bar\chi^2F+\chi\sigma^m\bar\chi\partial_m\varphi}{2F\overbar F-\tfrac{1}{2}\partial\varphi\partial\varphi}+\ldots~.
\end{equation}

{\bf The Lagrangian.} Let us review a general chiral superfield Lagrangian under the cubic constraint \cite{Aldabergenov:2021obf}. For simplicity we assume that the global shift-symmetry is exact, so that the axion $\varphi$ is massless, but it is always possible to add a small mass by breaking the shift-symmetry in K\"ahler potential or superpotential. Then the most general single-field, shift-symmetric model in global SUSY is given by
\begin{equation}\label{KW_gen_chiral}
    K=\tfrac{1}{2}(\bs+\overbar\bs)^2~,~~~W=\mu\bs~,
\end{equation}
where constant terms in $K$ and $W$, and linear term in $K$ are irrelevant (in global SUSY), and higher-order terms in $K$ are eliminated by the constraint $(\bs+\overbar\bs)^3=0$. In the superpotential, only the linear term in $\bs$ is allowed by the shift-symmetry. Eq. \eqref{KW_gen_chiral} leads to the Lagrangian
\begin{align}
\begin{aligned}
    \cl &=\int \!d^4\theta\, K+\left(\int\! d^2\theta\, W+{\rm h.c.}\right)\\
    &=\tfrac{1}{4}(\phi\Box\phi+\varphi\Box\varphi)-i\chi\sigma^m\partial_m\overbar\chi+\mu(F+\overbar F)+F\overbar F~,
\end{aligned}
\end{align}
where the cubic constraint eliminates $\phi$ according to Eqs. \eqref{phi_sol} and \eqref{phi_sol_leading}. The solution to the F-field equation of motion (EOM) has finite expansion in $\chi$ and $\bar\chi$ (after eliminating $\phi$), starting from the constant term, $F=-\mu+\co(\chi,\bar\chi)$ \cite{Aldabergenov:2021obf}. The resulting theory describes goldstino and axion dynamics with non-linear global $\cn=1$ SUSY.

\subsection{Nilpotent massive vector superfield}\label{sec_nil_vec_rigid}

The cubic constraint \eqref{cubic_constr} with global shift symmetry can be generalized for the case of local shift symmetry as well, which requires the introduction of an abelian gauge (vector) superfield $\bv$ with the component expansion
\begin{align}\label{V_hat_expansion}
\begin{aligned}
	\bv &=\alpha-\theta\sigma^m\bar\theta A_m+\tfrac{1}{2}\theta^2\bar\theta^2\big(D+\tfrac{1}{2}\Box\alpha\big)\\
	&+\Big[\sqrt{2}\theta\eta+\theta^2f+i\theta^2\bar\theta\big(\bar\lambda+\tfrac{1}{\sqrt{2}}\overbar\sigma^m\partial_m\bar\eta\big)+{\rm h.c.}\Big]~,
\end{aligned}
\end{align}
where $\alpha$ and $D$ are real scalars, $f$ is a complex scalar ($D$ and $f$ are auxiliary), $\eta$ and $\lambda$ are Weyl fermions, and $A_m$ is a real vector. If we choose the Wess--Zumino (WZ) gauge, the components $\alpha$, $\eta$, and $f$ vanish.

We then generalize the constraint as
\begin{equation}
	(\bs+\overbar\bs+\bv)^3=0~,\label{cubic_constr_SV}
\end{equation}
where the superfields transform as
\begin{equation}
	\bs\rightarrow\bs-\mathbf\Sigma~,~~~\bv\rightarrow\bv+\mathbf\Sigma+\overbar{\mathbf\Sigma}~,
\end{equation}
for some chiral superfield $\mathbf\Sigma$. Since shift symmetries are spontaneously broken everywhere in field space, we can work in the unitary gauge with $\mathbf\Sigma=\bs$, i.e. $\bv$ and $\bs$ combine into a single massive vector superfield. We denote this massive vector superfield as $\hat\bv$, and its vector component as $\hat A_m$, while for the rest of the components we keep the same notation as in \eqref{V_hat_expansion}.

The components $\alpha$, $\eta$, and $f$ combine with the components of $\bs$ (with ${\rm Re}(S)$, $\chi$, and $F$, respectively) and become physical. The pseudo-scalar ${\rm Im}(S)\equiv\varphi/2$ becomes the longitudinal mode of the massive vector
\begin{equation}
	\hat A_m=A_m+\partial_m\varphi~.
\end{equation}

In terms of the massive vector superfield $\hat\bv$, the constraint is simply
\begin{equation}
	\hat\bv^3=0~.\label{cubic_constr_V}
\end{equation}
Its highest component reads
\begin{equation}\label{cubic_constr_vec}
	\tfrac{1}{2}\alpha^2\big(D+\tfrac{1}{2}\Box\alpha\big)+\alpha\tilde B=\tilde H~,
\end{equation}
where
\begin{align}\label{BH_tilde_def}
\begin{aligned}
\tilde B &\equiv
	2f\bar f-\tfrac{1}{2}\hat A^m\hat A_m-\left(i\eta\sigma^m\partial_m\bar\eta-\sqrt{2}i\eta\lambda+{\rm h.c.}\right)~,\\
\tilde H &\equiv
	\eta^2\bar f+\bar\eta^2f+\eta\sigma^m\bar\eta\hat A_m~.
\end{aligned}
\end{align}
The constraint equation \eqref{cubic_constr_vec} is solved by
\begin{equation}\label{alpha_sol}
	\alpha=\frac{\tilde H}{\tilde B}-\frac{\tilde H^2}{2\tilde B^3}\left(D+\frac{\Box\tilde H}{2\tilde B}\right)=\frac{\eta^2\bar f+\bar\eta^2f+\eta\sigma^m\bar\eta\hat A_m}{2f\bar f-\tfrac{1}{2}\hat A^2}+\ldots~,
\end{equation}
where in the last equality we showed the leading-order (bilinear) solution in $\eta,\bar\eta$ (we denote $\hat A^2\equiv \hat A^m\hat A_m$). The solution is regular if $\tilde B$ is non-vanishing. Comparing this solution to the $\phi$-solution in the global shift symmetry case \eqref{phi_sol_leading}, it can be seen that the axion kinetic term $\partial\varphi\partial\varphi$ has been replaced by the vector mass term $\hat A^m\hat A_m$.

If the gauge multiplet is decoupled, we obtain the global $U(1)$ limit where the solution reduces to \eqref{phi_sol}. This can be seen by turning off the components of the original massless gauge multiplet, $\{\lambda,A_m,D\}\rightarrow 0$. In this case $\hat A_m\rightarrow \partial_m\varphi$, which restores the axion. After renaming
\begin{equation}
\alpha\rightarrow\phi~,~~~\eta\rightarrow\chi~,~~~f\rightarrow F~,
\end{equation}
the solution \eqref{alpha_sol} reduces to \eqref{phi_sol}.

{\bf The Lagrangian.} As for the chiral superfield case, we can try to write down the general globally supersymmetric Lagrangian for a single massive vector superfield satisfying $\hat\bv^3=0$ (and having up to two derivatives),
\begin{equation}\label{L_gen_vector}
    \cl=\int\!d^4\theta\left(\xi g\hat\bv+\tfrac{1}{2}g^2\hat\bv^2\right)-\left(\tfrac{1}{4}\int\!d^2\theta\,\boldsymbol\cw^\alpha \boldsymbol\cw_\alpha+{\rm h.c.}\right)~,
\end{equation}
where $\boldsymbol\cw_\alpha\equiv -\tfrac{1}{4}\overbar D^2D_\alpha\hat\bv$, $g$ is the gauge coupling, and $\xi$ is a real (Fayet--Iliopoulos) parameter. For normalization purposes we divide $\alpha$, $\eta$, and $f$ by $g$, and write the component Lagrangian as
\begin{align}
\begin{aligned}
    \cl=\tfrac{1}{4}(\alpha\Box\alpha-F_{mn}F^{mn}) &-\tfrac{i}{2}(\eta\sigma^m\partial_m\bar\eta+\lambda\sigma^m\partial_m\bar\lambda-\tfrac{1}{\sqrt{2}}g\eta\lambda+{\rm h.c.})\\
   & -\tfrac{1}{4}g^2\hat A_m\hat A^m+f\bar f+\tfrac{1}{2}g(\xi+\alpha)D+\tfrac{1}{2}g^2D^2~.
\end{aligned}
\end{align}
Here we run into a problem with using the constraint $\hat\bv^3=0$ and its solution \eqref{alpha_sol}: the $f$-field EOM becomes $f=0+\co(\chi,\bar\chi)$, i.e. it vanishes at the vacuum, which renders the solution \eqref{alpha_sol} singular (when $f=\hat A_m=0$). This means that the minimal model of a cubic nilpotent vector superfield in global $\cn=1$ SUSY is inconsistent at the vacuum (this is not the case in supergravity, as will be shown in the next section). However, it is possible to construct a consistent model if we add interactions with other superfields. For example we can introduce a chiral superfield $\Phi$ interacting with the vector $\hat\bv$ as
\begin{equation}
    \cl\supset\int\!d^4\theta\,\hat\bv(\Phi+\overbar\Phi)=\bar fF^{\Phi}+f\overbar F^{\Phi}+\ldots~,
\end{equation}
which modifies the $f$-EOM as $f=-F^\Phi+\co(\chi,\bar\chi)$. If $F^\Phi$ acquires a non-zero VEV, the model will be consistent with the cubic constraint (in this scenario, the goldstino will be a combination of $\eta$, the fermionic component of $\Phi$, and $\lambda$, if the D-field is also non-vanishing).

\section{Supergravity generalization}\label{sec_cubic_SUGRA}

\subsection{Constrained chiral superfield in SUGRA}

In supergravity we can use a curved superspace version of $\theta$-coordinate, denoted by $\Theta$, which is so defined that the chiral superfield has the same expansion as in rigid SUSY,
\begin{equation}
	\bs=S+\sqrt{2}\Theta\chi+\Theta^2F~.
\end{equation}
The Lagrangian for a chiral superfield then reads (we use the old-minimal formulation, see Appendices A and B for more details; when working in supergravity we set $M_{\rm Pl}=1$)
\begin{equation}\label{L_SG}
	\cl=\int\!d^2\Theta\,2\boldsymbol\ce\left[\tfrac{3}{8}(\overbar\cd^2-8\boldsymbol\car)e^{-K/3}+W\right]+{\rm h.c.}~,
\end{equation}
where $K=K(\bs,\overbar\bs)$ is the K\"ahler potential, and $W=W(\bs)$ is the superpotential. The components of density $\boldsymbol{\ce}$ and curvature $\boldsymbol\car$ superfields form the old-minimal ($12+12$) supergravity multiplet 
\begin{equation}
	\{e^a_m,\psi_m,M,b_m\}~,\nonumber
\end{equation}
where $e^a_m$ is the frame field, $\psi_m$ is gravitino (with suppressed spinor index), $M$ is complex auxiliary scalar, and $b_m$ is real auxiliary vector. $\cd_\alpha$~\footnote{We denote $\cd^2\equiv \cd^\alpha \cd_\alpha$ and $\overbar \cd^2\equiv\overbar\cd_{\dot\alpha}\overbar\cd^{\dot\alpha}$.} is the supergravity version of the superspace covariant derivative $D_\alpha$, and can be used to define the components of $\bs$,
\begin{equation}
	\bs|=S\equiv\tfrac{1}{2}(\phi+i\varphi)~,~~~\cd_\alpha\bs|=\sqrt{2}\chi_\alpha~,~~~\cd^2\bs|=-4F~.
\end{equation}
The mixed $\cd_\alpha$ and $\overbar\cd_{\dot\alpha}$ derivatives include in addition the supergravity multiplet contributions, shown in Appendix B, where a general structure of cubic constraints is also explained.

The supergravity version of the component form \eqref{cubic_constr_comp} of the cubic constraint can be obtained by applying $\cd^2\overbar\cd^2$ to the cubic constraint $(\bs+\overbar\bs)^3=0$. After some rearrangements we find,
\begin{equation}\label{cubic_constr_comp_SG}
    \tfrac{1}{2}\phi^2\left(2\ca+\tfrac{1}{2}\Box\phi\right)+\phi\left(\cb+\cc^m\partial_m\phi\right)=\ch~,
\end{equation}
where $\Box\equiv e^m_a\cd_m(e^a_n\cd^n)=\nabla_m\nabla^m$ is Laplace--Beltrami operator (see Appendix A for the definitions of covariant derivatives), and we use the following shorthands
\begin{align}\label{abch_def}
\begin{aligned}
	\ca &\equiv 
    \left(\tfrac{1}{3}MF+\tfrac{1}{4}\psi^m\psi_m F-\tfrac{1}{\sqrt{2}}\psi^m\cd_m\chi+{\rm h.c.}\right)+\tfrac{1}{3}\left(b^m-\tfrac{3}{4}\psi_n\sigma^m\bar\psi^n\right)\partial_m\varphi~,\\
	\cb &\equiv
	2F\overbar F-\tfrac{1}{2}\partial\varphi\partial\varphi-\tfrac{1}{3}\chi\sigma^m\bar\chi b_m-\chi\psi^m\bar\chi\bar\psi_m+\Big[-i\chi\sigma^m\cd_m\bar\chi+\tfrac{i}{\sqrt{2}}\chi\sigma^m\bar\psi_m\overbar F\\
	&\quad-\tfrac{3i}{2\sqrt{2}}\chi\psi^m\partial_m\varphi-\chi\sigma^{nm}\psi_m(\tfrac{i}{\sqrt{2}}\partial_n\varphi+\bar\chi\bar\psi_n)-\chi^2(\tfrac{1}{3}M+\tfrac{1}{4}\psi^m\psi_m)+{\rm h.c.}\Big]~,\\
	\cc^m &\equiv -\tfrac{1}{2\sqrt{2}}(\chi\psi^m-2\chi\sigma^{mn}\psi_n)+{\rm h.c.}~,\\
	\ch &\equiv
	\big[\chi^2\overbar F-\tfrac{i}{2\sqrt{2}}\chi^2\psi_m\sigma^m\bar\chi+{\rm h.c.}\big]+\chi\sigma^m\bar\chi\partial_m\varphi~.
\end{aligned}
\end{align}
The solution to \eqref{cubic_constr_comp_SG} is given by
\begin{equation}\label{phi_sol_SG}
	\phi=\frac{\ch}{\cb}-\frac{\ch}{\cb^3}\cc^m\partial_m\ch-\frac{\ch^2}{\cb^3}\left(\ca+\frac{\Box\ch}{4\cb}\right)~.
\end{equation}
If we turn off the supergravity multiplet as $\psi_m=M=b_m=0$ and $e^a_m=\delta^a_m$, this solution reduces to the solution \eqref{phi_sol} in the rigid SUSY limit since $\ca\rightarrow 0$, $\cb\rightarrow B$, $\cc_m\rightarrow 0$, and $\ch\rightarrow H$.

At the leading order from \eqref{phi_sol_SG} we again find
\begin{equation}\label{phi_sol_SG_leading}
	\phi=\frac{\chi^2\overbar F+\bar\chi^2F+\chi\sigma^m\bar\chi\partial_m\varphi}{2F\overbar F-\tfrac{1}{2}\partial\varphi\partial\varphi}+\ldots~.
\end{equation}
However, the component expansion of the supergravity Lagrangian \eqref{L_SG} is in Jordan frame, $\cl\supset e^{-K/3}\sqrt{-g}R/2$, and the Einstein frame is obtained after the Weyl rescaling,~\footnote{Under this Weyl rescaling the Pauli matrices with the curved index transform as $\sigma^m\rightarrow e^{-K/6}\sigma^m$, since $\sigma^m=e^m_a\sigma^a$.}
\begin{equation}\label{Weyl_resc}
	g_{mn}\rightarrow e^{K/3}g_{mn}~,~~~e^a_m\rightarrow e^{K/6}e^a_m~,
\end{equation}
while the ``canonical" fermion kinetic terms are obtained after the redefinitions
\begin{equation}\label{Fermi_resc}
	\chi\rightarrow e^{-K/12}\chi~,~~~\psi_m\rightarrow e^{K/12}\psi_m~,
\end{equation}
and the subsequent shift of the gravitino,
\begin{equation}\label{gravitino_shift}
	\psi_m\rightarrow\psi_m+\tfrac{\sqrt{2}i}{6}K_{\bar S}\sigma_m\bar\chi~.
\end{equation}
We can also rescale the $F$-field,
\begin{equation}\label{F_resc}
	F\rightarrow e^{-K/6}F~,
\end{equation}
so that its contribution to the scalar potential has the usual form $V\supset K_{S\overbar S}F\overbar F$. After applying the transformations \eqref{Weyl_resc}--\eqref{F_resc} we obtain an appropriately normalized Einstein frame action \cite{Wess:1992cp}. It is easy to check that the leading-order solution \eqref{phi_sol_SG_leading} to the cubic constraint is left invariant under these transformations, and coincides with the rigid SUSY case, while higher-order terms include contributions from the supergravity multiplet.

{\bf The Lagrangian.} Most general K\"ahler potential and superpotential for the chiral superfield $\bs$, satisfying the cubic constraint $(\bs+\overbar\bs)^3=0$ and invariant under global shifts $S\rightarrow S+ic$, read
\begin{equation}\label{KW_constrained_SUGRA}
    K=\xi(\bs+\overbar\bs)+\tfrac{1}{2}(\bs+\overbar\bs)^2~,~~~W=\mu~,
\end{equation}
where $\xi$ and $\mu$ are real parameters.
A constant term in the K\"ahler potential is irrelevant, since we can use the invariance of the Lagrangian under K\"ahler transformations (with a chiral superfield $\sigma$),
\begin{equation}\label{Kahler-Weyl}
    K\rightarrow K+\sigma+\overbar\sigma~,~~~W\rightarrow We^{-\sigma}~,
\end{equation}
to eliminate it by a constant rescaling of $W$. Unlike in global SUSY, in SUGRA, a constant term in the superpotential contributes to the Lagrangian, while the linear $\bs$-term is not allowed as it breaks the shift-symmetry. We can also use the transformation \eqref{Kahler-Weyl} to eliminate the linear term of the K\"ahler potential \eqref{KW_constrained_SUGRA} by choosing $\sigma=-\xi\bs$. With this choice, superpotential takes the form $W=\mu e^{\xi\bs}$, which is not invariant under the (imaginary) shift of $S$, but transforms by a constant phase. Thus, in this K\"ahler frame our shift-symmetry is identified with the global $U(1)_R$ symmetry. 

For convenience we use the K\"ahler frame \eqref{KW_constrained_SUGRA} where superpotential is constant. We then derive the relevant component action starting from the superfield Lagrangian \eqref{L_SG}, and by using the solution \eqref{phi_sol_SG} to the cubic constraint. The resulting Lagrangian is quite complicated, but it can be significantly simplified if we use the SUGRA unitary gauge where the goldstino, which is $\chi$ in this model, is set to zero. In this unitary gauge, the solution to the cubic constraint is simply $\phi=0$ (as before, we use $S=(\phi+i\varphi)/2$). Finally, in order to obtain the appropriately normalized Einstein frame action, we use the Weyl transformation \eqref{Weyl_resc}, and rescale the gravitino and the auxiliary field as:
\begin{equation}
    \psi\rightarrow e^{K/12}\psi~,~~~F\rightarrow e^{-K/6}F~.
\end{equation}
For $\phi=\chi=0$ (unitary gauge) this yields,
\begin{align}
\begin{aligned}
    e^{-1}\cl &=\tfrac{1}{2}R-\tfrac{1}{4}(K''-\tfrac{1}{3}{K'}^2)(\partial\varphi\partial\varphi-4F\overbar F)+\tfrac{1}{3}b_m(b^m+K'\partial^m\varphi)\\
    &-\mu e^{2K/3}(M+\overbar M)-\tfrac{1}{3}e^{K/6}K'(MF+\overbar M\overbar F)\\
    &-\left(\tfrac{1}{4}\epsilon^{klmn}\psi_k\sigma_l\bar\psi_{mn}+\mu e^{K/2}\psi_m\sigma^{mn}\psi_n+{\rm h.c.}\right)-\tfrac{i}{4}K'\epsilon^{klmn}\psi_k\sigma_l\bar\psi_m\partial_n\varphi~,
\end{aligned}
\end{align}
where $K'=\partial_SK=\partial_{\overbar S}K$ (since $K$ is a function of $S+\overbar S$). The next step is to eliminate the auxiliary fields by their EOM,
\begin{align}\label{aux_EOM}
\begin{aligned}
    b_m &=-\frac{K'}{2}\partial_m\varphi=-\frac{\xi}{2}\partial_m\varphi~,\\
    M &=\mu e^{K/3}\left(\frac{{K'}^2}{K''}-3\right)=\mu(\xi^2-3)~,\\
    F &=-\mu e^{K/2}\frac{K'}{K''}=-\mu\xi~,
\end{aligned}
\end{align}
where we used the fact that $K=0$, $K'=\xi$, and $K''=1$ when $\phi=0$. Plugging \eqref{aux_EOM} back into the Lagrangian, we find
\begin{align}\label{L_axion_gravitino}
\begin{aligned}
    e^{-1}\cl=\tfrac{1}{2}R-\tfrac{1}{4}\partial\varphi\partial\varphi &-\left(\tfrac{1}{4}\epsilon^{klmn}\psi_k\sigma_l\bar\psi_{mn}+\mu\psi_m\sigma^{mn}\psi_n+{\rm h.c.}\right)\\
    &-\tfrac{i}{4}\xi\epsilon^{klmn}\psi_k\sigma_l\bar\psi_m\partial_n\varphi-\mu^2(\xi^2-3)~.
\end{aligned}
\end{align}
Both $\mu$ and $\xi$ must be non-zero, otherwise the $F$-field vanishes, and the solution \eqref{phi_sol_SG_leading} becomes singular at the vacuum. Hence, the only consistent way to obtain Minkowski vacuum in this construction, is to set $\xi=\pm\sqrt{3}$. Non-zero $\xi$ also means that the gravitino-axion interaction, shown in the second line of \eqref{L_axion_gravitino}, is always present, while the model \eqref{KW_constrained_SUGRA} can always be K\"ahler-transformed to the frame where the shift symmetry is the R-symmetry ($W=\mu e^{\xi\bs}$), and $\varphi$ is its R-axion, regardless of the UV origin of the constrained superfield $\bs$.

The Lagrangian \eqref{L_axion_gravitino} represents the minimal model for the cubic constraint $(\bs+\overbar\bs)^3=0$, where the axion $\varphi$ is massless. However, a mass term, as well as other interactions in the Lagrangian, can be generated by various shift-symmetry-breaking terms in $K$ and $W$ (this is how e.g. minimal SUGRA inflationary models are constructed in this setup \cite{Terada:2021rtp,Aoki:2021nna}).

\subsection{Nilpotent massive vector superfield in SUGRA}\label{Subsec_nilpotent_V_Lagr}

Here we generalize the nilpotent massive vector superfield, $\hat\bv^3=0$, considered in Subsection \ref{sec_nil_vec_rigid} to supergravity. Applying supergravitational covariant derivatives $\cd^2\overbar\cd^2$ to the cubic constraint, and extracting the leading component yields~\footnote{The definitions of the components of $\hat\bv$ in supergravity can be found in Appendix B.}
\begin{equation}\label{nil_vec_comp_SG}
    \alpha^2\left(\tilde\ca+\tfrac{1}{4}\Box\alpha\right)+\alpha\left(\tilde\cb+\tilde\cc^m\partial_m\alpha\right)=\tilde\ch~,
\end{equation}
where
\begin{align}\label{abch_tilde_def}
\begin{aligned}
	\tilde\ca &\equiv\tfrac{1}{2}D+\left(\tfrac{1}{3}Mf+\tfrac{1}{4}\psi^m\psi_m f-\tfrac{1}{\sqrt{2}}\psi^m\cd_m\eta+{\rm h.c.}\right)+\tfrac{1}{3}\left(b^m-\tfrac{3}{4}\psi_n\sigma^m\bar\psi^n\right)\hat A_m~,\\
	\tilde\cb &\equiv
	2f\bar f-\tfrac{1}{2}\hat A^2-\tfrac{1}{3}\eta\sigma^m\bar\eta b_m-\eta\psi^m\bar\eta\bar\psi_m+\Big[-i\eta\sigma^m\cd_m\bar\eta+\sqrt{2}i\eta\lambda+\tfrac{i}{\sqrt{2}}\eta\sigma^m\bar\psi_m\bar f\\
	&\quad\qquad-\tfrac{3\sqrt{2}}{4}i\eta\psi^m\hat A_m-\eta\sigma^{nm}\psi_m(\tfrac{i}{\sqrt{2}}\hat A_n+\bar\eta\bar\psi_n)-\eta^2(\tfrac{1}{3}M+\tfrac{1}{4}\psi^m\psi_m)+{\rm h.c.}\Big]~,\\
	\tilde\cc^m &\equiv -\tfrac{1}{2\sqrt{2}}(\eta\psi^m-2\eta\sigma^{mn}\psi_n)+{\rm h.c.}~,\\
	\tilde\ch &\equiv
	\big[\eta^2\bar f-\tfrac{i}{2\sqrt{2}}\eta^2\psi_m\sigma^m\bar\eta+{\rm h.c.}\big]+\eta\sigma^m\bar\eta\hat A_m~.
\end{aligned}
\end{align}
This equation has the same basic form as Eq.\eqref{cubic_constr_comp_SG} for the global shift symmetry case, and the solution is
\begin{equation}\label{alpha_sol_SG}
	\alpha=\frac{\tilde\ch}{\tilde\cb}-\frac{\tilde\ch}{\tilde\cb^3}\tilde\cc^m\partial_m\tilde\ch-\frac{\tilde\ch^2}{\tilde\cb^3}\left(\tilde\ca+\frac{\Box\tilde\ch}{4\tilde\cb}\right)~.
\end{equation}

We can ``de-gauge" the shift symmetry by turning off $\{\lambda,A_m,D\}$. After renaming $\alpha\rightarrow\phi$ and $\eta\rightarrow\chi$, and using $\hat A_m\rightarrow\partial_m\varphi$, we recover the constraint for the chiral superfield \eqref{cubic_constr_comp_SG} and its solution \eqref{phi_sol_SG} with global shift symmetry. Another limit that we can take is the limit of global SUSY where $\psi_m=b_m=M=0$, which leads to the constraint \eqref{cubic_constr_vec}.

{\bf The Lagrangian.} A theory of a single massive vector multiplet is described by a real function $J(\hat\bv)$ (and its leading component $J(\alpha)$) which replaces the K\"ahler potential $K(\bs+\overbar\bs+\bv)$ in the unitary gauge where $\bs$ is set to zero. We start from the Lagrangian for a massive vector superfield with a general function $J(\hat\bv)$ in the presence of constant superpotential $\mu$,
\begin{equation}\label{L_SG_V}
	\cl=\int\!d^2\Theta\, 2\boldsymbol\ce\left[\tfrac{3}{8}(\overbar\cd^2-8\boldsymbol\car)e^{-J/3}+\mu+\tfrac{1}{4}\boldsymbol\cw^\alpha\boldsymbol\cw_\alpha\right]+{\rm h.c.}~.
\end{equation}
After extracting the components and performing field-dependend Weyl rescalings
\begin{gather}
\begin{gathered}
    e_m^a\rightarrow e^{J/6}e_m^a~,~~~\psi_m\rightarrow e^{J/12}\psi_m~,~~~\eta\rightarrow e^{-J/12}\eta~,\\
    \lambda\rightarrow e^{-J/4}\lambda~,~~~f\rightarrow e^{-J/6}f~,~~~D\rightarrow e^{-J/3}D~,
\end{gathered}
\end{gather}
and the gravitino shift~\footnote{See Appendix A for the transformation of the scalar curvature under these rescalings and the gravitino shift.}
\begin{equation}
    \psi^{\alpha}_m\rightarrow\psi^{\alpha}_m-\tfrac{\sqrt{2}}{6}iJ'\bar\eta_{\dot\alpha}\overbar\sigma^{\dot\alpha\alpha}_m~,
\end{equation}
we arrive at (we also restore the gauge coupling by dividing $\alpha,\eta,f$ by $2g$)
\begin{align}\label{L_SG_V_comp}
    e^{-1}\cl &=\tfrac{1}{2}R-\tfrac{1}{4}(J''\partial\alpha\partial\alpha+F_{mn}F^{mn})-g^2(J''-\tfrac{1}{3}{J'}^2)\hat A_m\hat A^m\nonumber\\
    &-\tfrac{i}{4}\epsilon^{klmn}\psi_k\sigma_l\bar\psi_m(2gJ'\hat A_n-J''\eta\sigma_n\bar\eta)-g(J'''-\tfrac{1}{3}J'J''+\tfrac{1}{18}{J'}^3)\eta\sigma^m\bar\eta\hat A_m\nonumber\\
    &-\tfrac{1}{2}(J''-\tfrac{1}{6}{J'}^2)\eta\psi_m\bar\eta\bar\psi^m+\tfrac{1}{12}{J'}^2\eta\lambda\bar\eta\bar\lambda+\tfrac{1}{2}\lambda\psi_m\bar\lambda\bar\psi^m+\tfrac{1}{4}\lambda\sigma^n\bar\psi^m\psi_n\sigma_m\bar\lambda\nonumber\\
    &+\tfrac{1}{4}(J^{(4)}-{J''}^2+\tfrac{5}{9}{J'}^2J''-\tfrac{4}{3}J'J'''-\tfrac{1}{18}{J'}^4)\eta^2\bar\eta^2-\tfrac{1}{3}e^{J/3}M\overbar M+(J''-\tfrac{1}{3}{J'}^2)f\bar f\nonumber\\
    &-\tfrac{1}{3}e^{J/6}J'(Mf+\overbar M\bar f)-\mu e^{2J/3}(M+\overbar M)+\tfrac{1}{6}e^{J/6}(J''-\tfrac{1}{3}{J'}^2)(\eta^2M+\bar\eta^2\overbar M)\nonumber\\
    &-\tfrac{1}{2}(J'''-J'J''+\tfrac{1}{9}{J'}^3)(\eta^2\bar f+\bar\eta^2f)+\tfrac{1}{3}b_mb^m+\tfrac{1}{2}\lambda\sigma^m\bar\lambda b_m-\tfrac{1}{6}(J''+\tfrac{1}{3}{J'}^2)\eta\sigma^m\bar\eta b_m\nonumber\\
    &+\tfrac{\sqrt{2}}{6}iJ'(\eta\psi^m-\bar\eta\bar\psi^m)b_m+\tfrac{2}{3}gJ'\hat A^mb_m+gJ'D+\tfrac{1}{2}D^2\nonumber\\
    &+\big\{-\tfrac{1}{4}\epsilon^{klmn}\psi_k\sigma_l\bar\psi_{mn}-\tfrac{i}{2}J''\eta\sigma^m\cd_m\bar\eta-\tfrac{i}{2}\lambda\sigma^m\cd_m\bar\lambda-\mu e^{J/2}\psi_m\sigma^{mn}\psi_n-\tfrac{1}{3}\mu e^{J/2}{J'}^2\eta^2\nonumber\\
    &+\sqrt{2}igJ''\eta\lambda-\tfrac{\sqrt{2}}{4}J''\eta\sigma^m\overbar\sigma^n\psi_m(\partial_n\alpha-2ig\hat A_n)+\tfrac{\sqrt{2}}{6}ig{J'}^2\eta\psi^m\hat A_m\nonumber\\
    &-\tfrac{i}{\sqrt{2}}\mu e^{J/2}J'\eta\sigma^m\bar\psi_m-\tfrac{1}{2}gJ'\lambda\sigma^m\bar\psi_m+\tfrac{i}{2}(F_{mn}-i\tilde F_{mn})\lambda\sigma^m\bar\psi^n+\tfrac{1}{48}{J'}^2\eta^2\psi_m\psi^m\nonumber\\
    &+\tfrac{\sqrt{2}}{48}i(J'J''+\tfrac{1}{3}{J'}^3)\eta^2\psi_m\sigma^m\bar\eta+\tfrac{\sqrt{2}}{8}iJ'\eta\psi_m\lambda\sigma^m\bar\lambda-\tfrac{1}{8}\lambda^2\bar\psi_m\overbar\sigma^{mn}\bar\psi_n\nonumber\\
    &-\tfrac{3}{16}\lambda^2\bar\psi_m\bar\psi^m+{\rm h.c.}\big\}~,
\end{align}
where $J^{(4)}$ means fourth derivative of $J$. Eq. \eqref{L_SG_V_comp} is a general (off-shell supersymmetric) Lagrangian for a single massive vector multiplet, equivalent to a theory of a chiral superfield with gauged shift-symmetry. If we eliminate the auxiliary fields, we reproduce the final SUGRA Lagrangian given in \cite{Wess:1992cp} for a particular case where the K\"ahler potential is a function of $\bs+\overbar\bs+\bv$. However, we are interested in the superfield constraint $\hat\bv^3=0$, whose solution is given by \eqref{alpha_sol_SG} and introduces additional fermionic terms with auxiliary field dependence. Unlike in the previous case with a single chiral superfield, here the goldstino is generally a combination of $\eta$ and $\lambda$, and therefore eliminating the goldstino (unitary gauge) does not significantly simplify the Lagrangian, while the solution to the cubic constraint cannot be set to zero. This must be taken into account before eliminating the auxiliary fields by their EOM.

Thanks to the cubic nilpotency of $\hat\bv$ and $\alpha$, we can write
\begin{equation}
    J(\alpha)=\xi\alpha+\tfrac{1}{2}\alpha^2~.
\end{equation}
A constant term in $J$ can always be absorbed by a redefinition of $\mu$ that parametrizes our constant superpotential. Before writing down the final Lagrangian for this model, we impose the supersymmetric unitary gauge,
\begin{equation}
    G=J'\left(\eta-\frac{\sqrt{2}g}{2\mu}ie^{-J/2}\lambda\right)=0~,
\end{equation}
where $G$ is the goldstino, in order to identify the physical fermion which we call $\zeta$. By using the solution \eqref{alpha_sol_SG} (where we apply the aforementioned Weyl rescalings and restore the gauge coupling), and subsequently eliminating the auxiliary fields, we obtain the Lagrangian,
\begin{align}\label{L_SG_V_comp_final}
\begin{aligned}
    e^{-1}\cl &=\frac{1}{2}R-\frac{1}{4}F_{mn}F^{mn}-g^2\hat A_m\hat A^m-\mu^2(\xi^2-3)-\frac{1}{2}g^2\xi^2\\
    &-\frac{i}{2}g\xi\epsilon^{klmn}\psi_k\sigma_l\bar\psi_m\hat A_n+\frac{g^3\xi}{2\mu^2+g^2}\left[\frac{1}{2}-\frac{\mu^2}{g^2}+\frac{1}{U}(\mu^2-g^2-\mu^2\xi^2)\right]\zeta\sigma^m\bar\zeta\hat A_m\\
    &+\Big\{-\frac{1}{4}\epsilon^{klmn}\psi_k\sigma_l\bar\psi_{mn}-\frac{i}{2}\zeta\sigma^m\cd_m\bar\zeta-\mu\psi_m\sigma^{mn}\psi_n\\
    &-\frac{\mu g^2}{4\mu^2+2g^2}\left[\xi^2-3+\frac{\xi^2}{U}(\mu^2-g^2-\mu^2\xi^2)\right]\zeta^2+\frac{ig^2\zeta\sigma^m\overbar\sigma^n\psi_m\hat A_n}{\sqrt{4\mu^2+2g^2}}\\
    &+\frac{\mu(F_{mn}-i\tilde F_{mn})}{\sqrt{4\mu^2+2g^2}}\zeta\sigma^m\bar\psi^n+{\rm h.c.}\Big\}+\text{4-fermion terms}~,
\end{aligned}
\end{align}
where $U\equiv\mu^2\xi^2-g^2\hat A_m\hat A^m$, and $\tilde F_{mn}\equiv\tfrac{1}{2}\epsilon_{mnkl}F^{kl}$. For $f$ and $D$ auxiliary fields we used the EOM,
\begin{equation}
    f=-\mu\xi+\ldots,~~~D=-g\xi+\ldots
\end{equation}
where $\ldots$ stands for $\zeta$-dependent terms. In the decoupling limit $g=0$ (and remembering that $\hat A_m=A_m+\tfrac{1}{2g}\partial_m\varphi$, in terms of the gauge field $A_m$ and the axion $\varphi$), we recover the Lagrangian \eqref{L_axion_gravitino} for the constrained chiral superfield plus a massless gauge multiplet. Therefore if we specifically want to describe a massive vector multiplet, we should assume $g\neq 0$. On the other hand, since consistency of the cubic constraint requires $\langle f\rangle=-\mu\xi\neq 0$, both $\mu$ and $\xi$ must be non-zero. This leads us to the point that both auxiliary fields ($f$ and $D$) of the massive vector superfield, subject to the cubic constraint $\hat\bv^3=0$, are non-vanishing.

As for the masses of the physical component fields, we have (in SUGRA Planck units) $m_{\hat A}=\sqrt{2}g$ for the vector field, and $m_\zeta=g^2/\mu$ for the spin-1/2 fermion, where we used $\langle U\rangle=\mu^2\xi^2$. The gravitino mass is $m_{3/2}=\mu$ as before. We have no scalar fields left, since $\alpha$ is eliminated by the constraint, and the cosmological constant is given by
\begin{equation}
    V=\mu^2(\xi^2-3)+\tfrac{1}{2}g^2\xi^2~.
\end{equation}
In Minkowski vacuum we can fix
\begin{equation}
    \xi^2=\frac{6}{2+g^2/\mu^2}~.
\end{equation}

\section{UV models for the cubic constraints}\label{sec_UV}

Here we consider several microscopic models (in supergravity) with spontaneously broken abelian symmetry and SUSY, which lead to the cubic constraints for chiral or vector (when the abelian symmetry is gauged) superfields.

\subsection{A global SUSY model}

Let us start from global SUSY for simplicity. A (globally) shift-symmetric UV model that can reproduce the constraint $(\mathbf S+\overbar{\mathbf S})^3=0$ was proposed in Ref. \cite{Aoki:2021nna}, defined by the following K\"ahler potential and superpotential,
\begin{equation}\label{KW_global_shift}
	K=\frac{1}{2}(\mathbf S+\overbar{\mathbf S})^2-\frac{1}{24\Lambda^2}(\mathbf S+\overbar{\mathbf S})^4~,~~~W=\mu\mathbf S~,
\end{equation}
with real parameters $\Lambda$ (with units of mass) and $\mu$ (with units of mass-squared). The model has the $U(1)$ symmetry which shifts the scalar $S$,
\begin{equation}
	S\rightarrow S+ic~,
\end{equation}
with a real parameter $c$. The superpotential shifts by an irrelevant (in global SUSY) constant, which does not explicitly break the $U(1)$ shift symmetry. Let us now parametrize $S=(\phi+i\varphi)/2$ (for almost canonical kinetic terms). The scalar potential becomes
\begin{equation}
	V_F=K^{\bar SS}W_S\overbar W_{\bar S}=\mu^2\left(1-\frac{\phi^2}{2\Lambda^2}\right)^{-1}~,
\end{equation}
which is minimized for $\langle\phi\rangle=0$. The canonical $\phi$ mass squared is given by $m^2_\phi=2\mu^2/\Lambda^2$ (proportional to the fourth derivative of the K\"ahler potential), and thus $\phi$ can be decoupled at the formal limit $\Lambda\rightarrow 0$.~\footnote{There is a subtlety related to taking the limit of small $\Lambda$ due to the fact that the higher-order terms, such as the quartic term in $K$, often originate from integrating out a heavy field(s) in some UV completion of the model, with $\Lambda$ being related to its mass. If $\Lambda$ is too small, this may destabilize some heavy field from its minimum. This situation was studied in Ref. \cite{Dudas:2016eej} (in relation to quadratic superfield constraints), where it was shown that there can exist a parameter range in the UV theory, where $\Lambda$ is small enough to describe the constrained superfield, but not too small as to destabilize the heavy fields.} This limit imposes our cubic constraint as can be seen from the superfield equation of motion,
\begin{equation}\label{Super_EOM_rigid}
	D^2\bs-\frac{1}{3\Lambda^2}D^2(\bs+\overbar\bs)^3-4\mu=0~.
\end{equation}
The leading term of the above equation gives the F-term EOM,
\begin{equation}
	F=-K^{\bar SS}\left[\mu-\frac{\chi^2}{2\Lambda^2}(S+\overbar S)\right]~,
\end{equation}
where at the vacuum we have $\langle F\rangle=-\mu$ spontaneously breaking SUSY, while $\chi$ plays the role of the goldstino. Now, if we take $\Lambda\rightarrow 0$ (while keeping $\mu$ finite), it can be seen that $(\bs+\overbar\bs)^3=0$ is a particular solution to the superfield equation \eqref{Super_EOM_rigid}. To see that it is also a unique (non-trivial) solution we check the equation of motion for $\phi$, which can be obtained by applying $\overbar D^2$ to \eqref{Super_EOM_rigid}. The real part of the resulting equation in the limit $\Lambda\rightarrow 0$ reads
\begin{equation}\label{phi_EOM_rigid}
	\phi(\tfrac{1}{2}\phi\Box\phi+\tfrac{1}{2}\partial\phi\partial\phi+B)=H~,\end{equation}
where $B$ and $H$ are defined in \eqref{BH_def}. $H$ is bilinear in $\chi$ and $\bar\chi$, so that $H^3=0$, which leads to $\phi^3=0$ by looking at the above equation (the cube of the expression in the parentheses cannot identically vanish since $B=2F\overbar F+\ldots$ and $F\propto \mu\neq 0$). Then by using $\Box\phi^3=0$, Eq. \eqref{phi_EOM_rigid} can be rewritten as
\begin{equation}
	\tfrac{1}{4}\phi^2\Box\phi+\phi B=H~,
\end{equation}
which is exactly our cubic constraint in the component form \eqref{cubic_constr_comp}. Its solution is given by \eqref{phi_sol} and \eqref{phi_sol_leading} where $\phi$ is eliminated in terms of bilinear and higher-order terms in $\chi,\bar\chi$, thus describing a low-energy effective theory of the axion $\varphi$ and the goldstino $\chi$ with non-linearly realized $\cn=1$ SUSY.

\subsection{R-symmetric model in supergravity}

In supergravity, the linear superpotential as in \eqref{KW_global_shift} breaks the shift symmetry of $S$, but we can instead introduce an exponential dependence,
\begin{equation}\label{UV_model_SUGRA_R}
	K=\frac{1}{2}(\mathbf S+\overbar{\mathbf  S})^2-\frac{1}{24\Lambda^2}(\mathbf S+\overbar{ \mathbf S})^4~,~~~W=\mu e^{\xi \mathbf S}~,
\end{equation}
where $\Lambda,\mu,\xi$ are real parameters. If $\xi\neq 0$, the global shift symmetry ($S\rightarrow S+ic$) is an R-symmetry. For convenience, we K\"ahler-transform \eqref{UV_model_SUGRA_R} into
\begin{equation}\label{UV_model_SUGRA}
	K=\xi(\mathbf S+\overbar{\mathbf  S})+\frac{1}{2}(\mathbf S+\overbar{\mathbf  S})^2-\frac{1}{24\Lambda^2}(\mathbf S+\overbar{\mathbf  S})^4~,~~~W=\mu~,
\end{equation}
so that the superpotential is constant. Since this model is equivalent to \eqref{UV_model_SUGRA_R} (assuming $\xi\neq 0$), we will still refer to the shift symmetry of $S$ as the R-symmetry.

We again parametrize $S=(\phi+i\varphi)/2$, and write the F-term scalar potential as
\begin{gather}\label{V_F_global_shift}
\begin{gathered}
	V_F=e^K\left(K^{\bar SS}D_SWD_{\bar S}\overbar W-3W\overbar W\right)\\
	=\mu^2\exp\left(\xi\phi+\frac{\phi^2}{2}-\frac{\phi^4}{24\Lambda^2}\right)\left[\left(1-\frac{\phi^2}{2\Lambda^2}\right)^{-1}\left(\xi+\phi-\frac{\phi^3}{6\Lambda^2}\right)^2-3\right]~.
\end{gathered}
\end{gather}
The corresponding vacuum equations are hard to solve exactly, but we can try to solve them perturbatively for small $\Lambda$. However, if $\Lambda$ is small and $\langle S\rangle$ is non-zero (as a general assumption), the quartic correction to the canonical K\"ahler potential may become large and spoil its perturbative form. Therefore let us assume that $S$, or rather $|\phi|$, is always smaller than $\Lambda$, which is encoded in the parametrization
\begin{equation}\label{phi_hat_par}
	\phi=\Lambda^2\hat\phi~.
\end{equation}
Using \eqref{phi_hat_par} and taking the series expansion of the potential \eqref{V_F_global_shift} in small $\Lambda$, we write down the Minkowski vacuum equations,
\begin{align}\label{VE_global_shift}
\begin{aligned}
	V_F/\mu^2 &=\xi^2-3+\xi(\xi^2-1)\Lambda^2\hat\phi+\tfrac{1}{2}\xi^2\Lambda^2\hat\phi^2+\co(\Lambda^4)=0~,\\
	\partial_{\hat\phi}V_F/\mu^2 &=\xi(\xi^2-1)\Lambda^2+\xi^2\Lambda^2\hat\phi+\co(\Lambda^4)=0~.
\end{aligned}
\end{align}
The vacuum equations are solved by
\begin{equation}\label{Vsol_global_shift}
	\xi^2=3+2\Lambda^2+\co(\Lambda^4)~,~~~\hat\phi=\frac{1}{\xi}-\xi+\co(\Lambda^2)\approx -\frac{2}{\sqrt{3}}~,
\end{equation}
where we have chosen the ``plus" branch, $\xi=+\sqrt{3}$, i.e. the VEV of $\phi$ becomes $\langle\phi\rangle\simeq -2\Lambda^2/\sqrt{3}$, which is consistent with our assumptions. To confirm these estimates we also numerically solved the vacuum equations. For $\Lambda=0.1$ the perturbative solution \eqref{Vsol_global_shift} yields $\xi=1.73781$ and $\langle\phi\rangle=-1.15\times 10^{-2}$, while from the numerical calculation we have $\xi=1.73779$ and $\langle\phi\rangle=-1.14\times 10^{-2}$. For $\Lambda=0.01$ both perturbative and numerical solutions yield $\xi=1.73211$ and $\langle\phi\rangle=-1.15\times 10^{-4}$. These results show that at the limit $\Lambda\rightarrow 0$, the VEV $\langle\phi\rangle$ goes to zero as $\Lambda^2$ and can be ignored for our purposes, while the parameter $\xi$ must be close to $\pm\sqrt{3}$ in order to cancel the cosmological constant. At the vacuum, the K\"ahler metric is (nearly) canonical, $\langle K_{S\bar S}\rangle=1+\co(\Lambda^2)$.

The leading-order expressions for the (canonical) mass of $\phi$, the $F$-field and the gravitino mass at the vacuum are given by~\footnote{Before writing the auxiliary field VEVs, we always use their conventional expressions, $F=-e^{K/2}K^{S\bar S}D_{\bar S}\overbar W$, $D=-g\mathscr{D}$, where $\mathscr{D}$ is Killing potential given by \eqref{Killing_pot}. This requires the rescaling $F\rightarrow e^{-K/6}F$ and $D\rightarrow e^{-K/3}D$, since in curved superspace formalism we start in Jordan frame.}
\begin{equation}\label{global_R_masses}
	m_\phi/\mu=\sqrt{6}/\Lambda+\co(\Lambda^0)~,~~~\langle m_{3/2}\rangle/\mu=1+\co(\Lambda^2)~,~~~\langle F\rangle/\mu=-\sqrt{3}+\co(\Lambda^2)~,
\end{equation}
assuming positive $\mu$ and $\Lambda$.
Similarly to the globally supersymmetric model \eqref{KW_global_shift} of the previous subsection, the $\phi$-mass diverges when $\Lambda\rightarrow 0$, while SUSY breaking scale remains finite, which suggests that the cubic constraint can be used to describe the system in the infrared.

To show this we derive superfield EOM for this model ($K$ and $W$ given by \eqref{UV_model_SUGRA}),
\begin{equation}\label{global_R_super_EOM}
	(\overbar\cd^2-8\boldsymbol\car)(e^{-K/3}K_\bs)=0~,
\end{equation}
When $\Lambda\rightarrow 0$ we have $K_{\bs}\simeq -(\bs+\overbar\bs)^3/(6\Lambda^2)$, which means $(\bs+\overbar\bs)^3=0$ solves \eqref{global_R_super_EOM} in this formal limit. To prove that it is a unique solution, we will solve \eqref{global_R_super_EOM} in components. In particular, applying $\cd^2$ to this equation (after multiplying by $e^{K/3}$ for convenience), we obtain the equation of motion for $\phi$, which in the limit $\Lambda\rightarrow 0$ reads
\begin{align}\label{global_R_EOM_limit}
\begin{aligned}
	-\ch &+\phi\left(\cb+\tfrac{1}{2}\partial\phi\partial\phi+\cc^m\partial_m\phi+\tfrac{1}{6}\chi^2M+\tfrac{1}{6}\bar\chi^2\overbar M\right)\\
	&+\tfrac{1}{2}\phi^2\left(2\ca+\Box\phi-\tfrac{1}{3}MF-\tfrac{1}{3}\overbar M\overbar F\right)+\co(\phi^3)=0~,
\end{aligned}
\end{align}
where $\ca,\cb,\cc^m,\ch$ are given by \eqref{abch_def}, below we repeat these for convenience,
\begin{align}\label{abch_def2}
\begin{aligned}
	\ca &\equiv 
    \left(\tfrac{1}{3}MF+\tfrac{1}{4}\psi^m\psi_m F-\tfrac{1}{\sqrt{2}}\psi^m\cd_m\chi+{\rm h.c.}\right)+\tfrac{1}{3}\left(b^m-\tfrac{3}{4}\psi_n\sigma^m\bar\psi^n\right)\partial_m\varphi~,\\
	\cb &\equiv
	2F\overbar F-\tfrac{1}{2}\partial\varphi\partial\varphi-\tfrac{1}{3}\chi\sigma^m\bar\chi b_m-\chi\psi^m\bar\chi\bar\psi_m-\Big[i\chi\sigma^m\cd_m\bar\chi-\tfrac{i}{\sqrt{2}}\chi\sigma^m\bar\psi_m\overbar F\\
	&\quad+\tfrac{3i}{2\sqrt{2}}\chi\psi^m\partial_m\varphi+\chi\sigma^{nm}\psi_m(\tfrac{i}{\sqrt{2}}\partial_n\varphi+\bar\chi\bar\psi_n)+\chi^2(\tfrac{1}{3}M+\tfrac{1}{4}\psi^m\psi_m)+{\rm h.c.}\Big]~,\\
	\cc^m &\equiv -\tfrac{1}{2\sqrt{2}}(\chi\psi^m-2\chi\sigma^{mn}\psi_n)+{\rm h.c.}~,\\
	\ch &\equiv
	\big[\chi^2\overbar F-\tfrac{i}{2\sqrt{2}}\chi^2\psi_m\sigma^m\bar\chi+{\rm h.c.}\big]+\chi\sigma^m\bar\chi\partial_m\varphi~.
\end{aligned}
\end{align}
It is immediately clear from \eqref{global_R_EOM_limit} that $\phi$ is proportional to $\ch$ (assuming $\cb$ is non-vanishing at the vacuum) which is at least bilinear in $\chi$ and $\bar\chi$, and this means that $\phi^3=0$. We then rewrite \eqref{global_R_EOM_limit} as
\begin{equation}\label{global_R_EOM_2}
	\tfrac{1}{2}\phi^2\left(2\ca+\tfrac{1}{2}\Box\phi\right)+\phi\left(\cb+\cc^m\partial_m\phi\right)-\ch=\tfrac{1}{6}M\left(\phi^2F-\phi\chi^2\right)+{\rm h.c.}~,
\end{equation}
where the right hand side (RHS) is the difference of this equation from the component form \eqref{cubic_constr_comp_SG} of the cubic constraint $(\bs+\overbar\bs)^3=0$. It can be shown however, that the RHS vanishes as follows. We first notice that the expression $\phi^2F-\phi\chi^2$ is fourth order in $\chi,\bar\chi$ (since $\phi$ is proportional to $\ch$ which is at least bilinear). So in order to find its explicit form, it is enough to find the solution for $\phi$ up to bilinear terms (since higher-order terms will vanish upon substitution). The leading-order (bilinear) solution to \eqref{global_R_EOM_2} is $\phi=\ch/\cb$, and it is then a simple exercise to use the definitions of $\ch$ and $\cb$ from \eqref{abch_def2} and show that
\begin{equation}
	\phi^2F-\phi\chi^2=0~.
\end{equation}
That is, the RHS of \eqref{global_R_EOM_2} vanishes and we obtain exactly the cubic constraint \eqref{cubic_constr_comp_SG}. EOM \eqref{global_R_super_EOM} is of course in Jordan frame, and the Einstein frame can be obtained by using transformations \eqref{Weyl_resc}--\eqref{F_resc}.

\subsection{Locally R-symmetric model and nilpotent vector superfield}\label{subsec_local_R_UV}

We now gauge the R-symmetry of the previous model, and derive the cubic constraint on the massive vector superfield from its equation of motion.

After including the $U(1)_R$ gauge multiplet, the superfield Lagrangian can be written as
\begin{equation}\label{gauge_R_super_L}
    {\cal L}=\int\! d^2\Theta\, 2\boldsymbol{\cal E}\left[\tfrac{3}{8}(\overbar{\cal D}^2-8\boldsymbol{\cal R})e^{-K/3}+\mu+\tfrac{1}{4}\boldsymbol{\cal W}^\alpha\boldsymbol{\cal W}_\alpha\right]+{\rm h.c.}~,
\end{equation}
where we have
\begin{equation}\label{gauge_R_model}
	K=\xi(\bs+\overbar\bs+\bv)+\frac{1}{2}(\bs+\overbar\bs+\bv)^2-\frac{1}{24\Lambda^2}(\bs+\overbar\bs+\bv)^4~,~~~W=\mu~.
\end{equation}
Next, we choose the superfield unitary gauge,
\begin{equation}
    \bs+\overbar\bs+\bv\rightarrow\hat\bv~,
\end{equation}
for a massive vector superfield $\hat\bv$, and the K\"ahler potential becomes
\begin{equation}\label{gauge_R_J}
    K(\bs+\overbar\bs+\bv)\rightarrow J(\hat\bv)=\xi\hat\bv+\frac{\hat\bv^2}{2}-\frac{\hat\bv^4}{24\Lambda^2}~,
\end{equation}
and we denote it as $J(\hat\bv)|$. From the Lagrangian \eqref{gauge_R_super_L} by using \eqref{gauge_R_J} we can derive the equation of motion for the massive vector superfield,
\begin{equation}
    e^{-J/3}J'=\cd^\alpha\boldsymbol{\cw}_\alpha~,
\end{equation}
where we used the reality property $\cd^\alpha \boldsymbol{\cw}_\alpha=\overbar\cd_{\dot\alpha}\overbar {\boldsymbol{\cw}}^{\dot\alpha}$. The leading component of the above equation eliminates the auxiliary D-field, while extracting its $\overbar\cd^2\cd^2$-component yields EOM for the scalar field $\alpha=\hat\bv|$ in Jordan frame,
\begin{align}
\begin{aligned}\label{gauge_R_EOM}
    4Y_1\big\{\tfrac{1}{2}\Box\alpha+\tilde\ca+\big[-\tfrac{1}{2\sqrt{2}}\eta \cd_m\psi^m-\tfrac{i}{4\sqrt{2}}\eta\psi_m(\psi_n\sigma^m\overbar\psi^n+\tfrac{1}{3}b^m)\\
    -\tfrac{\sqrt{2}}{12}\eta\sigma^{mn}(\psi_{mn}+i\psi_mb_n)+{\rm h.c.}\big]\big\}+Y_2(\partial\alpha\partial\alpha+2\tilde\cc^m\partial_m\alpha+2\tilde\cb)\\
    -2Y_3\tilde\ch+Y_4\eta^2\bar\eta^2-\tfrac{1}{8}e^{J/3}(\overbar\cd^2\cd^2\overbar\cd_{\dot\alpha}\overbar{\boldsymbol{\cw}}^{\dot\alpha}|+{\rm h.c.})=0~,
\end{aligned}
\end{align}
where we multiplied the equation by $e^{J/3}$, and used the notation
\begin{align}
\begin{aligned}
    Y_1 &\equiv J''-\tfrac{1}{3}{J'}^2~,\\
    Y_{1+i} &\equiv Y_i'-\tfrac{1}{3}J'Y_i~,
\end{aligned}
\end{align}
for $i=1,2,3$. The functions $\tilde\ca,\tilde\cb,\tilde\cc_m,\tilde\ch$ are defined in Eq. \eqref{abch_tilde_def}. The last term of \eqref{gauge_R_EOM}, $\overbar\cd^2\cd^2\overbar\cd\overbar{\boldsymbol{\cw}}|$, involves the vector multiplet and supergravity multiplet components, but we will not need its explicit form.

Before studying the EOM, let us comment on the scalar potential and vacuum structure of this model comparing it to the previous model with global $U(1)_R$. Here we gauged the $U(1)_R$ and chose a K\"ahler frame where the superpotential is just a constant $W=\mu$, which leads to the following scalar potential in the Einstein frame,
\begin{equation}\label{gauge_R_V}
    V=\mu^2e^J\left(\frac{{J'}^2}{J''}-3\right)+\frac{g^2}{2}{J'}^2~,
\end{equation}
after restoring the gauge coupling by $\alpha\rightarrow\alpha/(2g)$. The bosonic sector of the full Lagrangian is given by
\begin{equation}
    e^{-1}\cl_{\rm bos}=\tfrac{1}{2}R-\tfrac{1}{4}(J''\partial\alpha\partial\alpha+F_{mn}F^{mn})-g^2J''\hat A_m\hat A^m-V~,
\end{equation}
where $F_{mn}$ is the field strength of $\hat A_m$.

As in the previous model, we assume that $\langle\alpha\rangle$ is of the order $\Lambda^2$. Perturbatively solving the vacuum equations $V=V'=0$ in small $\Lambda$, we obtain
\begin{equation}
    \xi^2=\frac{6}{2+\hat g^2}+\frac{(4+\hat g^2+\hat g^4)^2}{(2+\hat g^2)^3}\Lambda^2+\co(\Lambda^4)~,~~~\langle\alpha\rangle=\frac{1}{\xi}(1-\xi^2-\hat g^2)\Lambda^2+\co(\Lambda^4)~,
\end{equation}
where $\hat g\equiv g/\mu$. When $g=0$ and $\alpha=\phi$ this agrees with the previous model where the solution is given by \eqref{Vsol_global_shift}. The canonical mass spectrum at the Minkowski vacuum reads
\begin{gather}\label{UV_gauge_R_masses}
\begin{gathered}
    m_\alpha=\frac{2\sqrt{3}\mu}{\sqrt{2\mu^2+g^2}\Lambda}+\co(\Lambda^0)~,~~~m_{\hat A}=\sqrt{2}g+\co(\Lambda^2)~,\\
    m_{1/2}=g^2/\mu+\co(\Lambda^2)~,~~~m_{3/2}=\mu+\co(\Lambda^2)~,\\
    \langle f\rangle=-\frac{\sqrt{6}\mu^2}{\sqrt{2\mu^2+g^2}}+\co(\Lambda^2)~,~~~\langle D\rangle=-\frac{\sqrt{6}\mu g}{\sqrt{2\mu^2+g^2}}+\co(\Lambda^2)~,
\end{gathered}
\end{gather}
where $m_{1/2}$ is the mass of the physical fermion orthogonal to the goldstino. The scalar mass $m_\alpha$ diverges as $1/\Lambda$ (as before), while the rest of the masses and SUSY breaking parameters are finite, as long as $g/\mu$ stays finite. By gauging the abelian symmetry, we obtained a massive vector, massive spinor, and a non-vanishing D-field, compared to the previous model with a single chiral superfield.

Going back to the $\alpha$-EOM \eqref{gauge_R_EOM}, in the limit $\Lambda\rightarrow 0$ we can approximate the $Y$-functions as
\begin{equation}
    Y_1\simeq -\frac{\alpha^2}{2\Lambda^2}+\co(\alpha^6)~,~~~Y_2\simeq -\frac{\alpha}{\Lambda^2}+\co(\alpha^5)~,~~~Y_3\simeq -\frac{1}{\Lambda^2}+\co(\alpha^4)~,~~~Y_4\simeq \co(\alpha^3)~.
\end{equation}
Thus, the $Y_4$-term and the last term of \eqref{gauge_R_EOM} (proportional to $e^{J/3}$) are at least cubic in $\alpha$ in this limit. It can then be shown from \eqref{gauge_R_EOM} that $\alpha$ is proportional to $\tilde\ch$, which is at least bilinear in $\eta$ and $\bar\eta$, which means that $\alpha^3$ identically vanishes (as well as $\alpha^2\eta$ and $\alpha^2\bar\eta$). The resulting equation for $\alpha$ coincides with Eq. \eqref{nil_vec_comp_SG} describing the cubic nilpotent massive vector superfield, while the mass spectrum \eqref{UV_gauge_R_masses} (when $\Lambda\rightarrow 0$) matches the mass spectrum of the Lagrangian \eqref{L_SG_V_comp_final}, as expected.

\subsection{Heavy dilaton model}

Here we consider a toy model for the string dilaton,
\begin{equation}\label{dilaton_KW}
    K=-\log(\bs+\overbar\bs+k)~,~~~W=\mu~,
\end{equation}
where the real scalar $\phi\equiv S+\overbar S$ is the dilaton, the imaginary part $\varphi\equiv -i(S-\overbar S)$ is the axion, $\mu$ is a real constant (which is arbitrary but non-zero) as before, and we assume the invariance under global shifts, $S\rightarrow S+ic$. $k$ is a function of $\bs+\overbar\bs$, and denotes (perturbative or non-perturbative) quantum corrections to the tree-level dilaton K\"ahler potential. Since the dilaton VEV determines the string coupling as $\phi\propto 1/g_s^2$, we require that the correction term $k$ vanishes in the weak coupling limit $\phi\rightarrow\infty$.

In order to identify in which case the model can be described by the cubic constraint, we can look at the superfield EOM, which for constant superpotential reads
\begin{equation}\label{dilaton_super_EOM}
	(\overbar\cd^2-8\boldsymbol{\car})(e^{-K/3}K')=0~,
\end{equation}
where $K'=K_\bs=K_{\overbar\bs}$. We expand it around the vacuum as
\begin{equation}\label{Kp_series}
    K'=\langle K'\rangle+\langle K''\rangle\Omega+\tfrac{1}{2}\langle K'''\rangle\Omega^2+\tfrac{1}{6}\langle K^{(4)}\rangle\Omega^3+\ldots~,
\end{equation}
where $\Omega\equiv\bs+\overbar\bs-\langle \phi\rangle$, and $K^{(4)}$ denotes the fourth derivative of the K\"ahler potential. In the previous model \eqref{UV_model_SUGRA} we had $\langle K^{(4)}\rangle\rightarrow\infty$ (and $\langle\phi\rangle\rightarrow 0$), as $\Lambda\rightarrow 0$, while the first three terms in the expansion \eqref{Kp_series} of its K\"ahler potential were finite. As was shown, this leads to the effective constraint $(\bs+\overbar\bs)^3=0$, and since the mass of $\phi$ includes the fourth derivative of $K$, it also diverges in that model for $\Lambda\rightarrow 0$. 

Here we show an alternative route to the cubic constraint, where $\langle K^{(4)}\rangle$ remains finite, and instead, $\langle K'\rangle$, $\langle K''\rangle$, and $\langle K'''\rangle$ go to zero in an appropriate limit. In this case we still have a diverging $\phi$-mass, since in the canonical normalization it is equal to $m_{\phi}=\sqrt{2\langle V''\rangle/\langle K''\rangle}$ ($V'=\partial_\phi V$). Therefore it is convenient to introduce a small parameter $\delta$ as,
\begin{equation}\label{Kpp_delta}
    \langle K''\rangle\equiv\delta^2~,
\end{equation}
such that in the limit $\delta\rightarrow 0$, the canonical mass $m_\phi$ goes to infinity (provided that $\langle V''\rangle$ is non-vanishing). The Minkowski vacuum condition in this case reads
\begin{equation}\label{V_eq_dilaton}
    V/\mu^2=e^K\left(\frac{{K'}^2}{K''}-3\right)=e^K\left(\frac{{K'}^2}{\delta^2}-3\right)=0~.
\end{equation}
Therefore we have the condition
\begin{equation}\label{Kp_delta}
    \langle K'\rangle=\sqrt{3}\delta~,
\end{equation}
which fits our requirement of small $\langle K'\rangle$. We also have the stationary point equation which, by using \eqref{V_eq_dilaton} can be written as
\begin{equation}
    V'/\mu^2=e^KK'\left(2-\frac{K'K'''}{{K''}^2}\right)=0~.
\end{equation}
After using \eqref{Kpp_delta} and \eqref{Kp_delta}, this leads to
\begin{equation}\label{Kppp_delta}
    \langle K'''\rangle=\tfrac{2}{\sqrt{3}}\delta^3~,
\end{equation}
which, together with \eqref{Kpp_delta} and \eqref{Kp_delta}, must be satisfied if we want to obtain the cubic constraint while having stable Minkowski vacuum (additionally, $\langle K^{(4)}\rangle$ must be non-vanishing when $\delta\rightarrow 0$). Under these vacuum conditions, the canonical dilaton mass becomes
\begin{equation}
    m_\phi=\sqrt{\frac{2\langle V''\rangle}{\langle K''\rangle}}=\sqrt{6}\mu e^{\langle K\rangle/2}\left(-\frac{\langle K^{(4)}\rangle}{\delta^4}+2\right)^{1/2}~,
\end{equation}
and the VEV of $F$ is
\begin{equation}
    \langle F\rangle=-\sqrt{3}\mu e^{\langle K\rangle/2}/\delta~.
\end{equation}
Assuming that $\langle K\rangle$ is finite, $\langle F\rangle$ diverges as $1/\delta$. However its contribution to the Lagrangian, $\langle K''F\overbar F\rangle$, is finite because $\langle K''\rangle=\delta^2$. In fact, the chiral superfield $\bs$ should be rescaled by $1/\delta$ in order to obtain canonically normalized effective K\"ahler metric, as we will do below after considering a specific example.

Next, we search for a K\"ahler potential for the dilaton-axion superfield that meets our conditions. At this point it is desirable to be able to control the dilaton VEV, in order to make sure the vacuum is in the perturbative regime in terms of $1/\phi$. Therefore we treat $\langle\phi\rangle$ as a free parameter which we can fix by hand, and introduce three parameters in the correction term $k$, in order to satisfy the three equations \eqref{Kpp_delta}, \eqref{Kp_delta}, and \eqref{Kppp_delta}. We find that various combinations of perturbative (in $1/\phi$) and non-perturbative corrections can satisfy the aforementioned conditions. Perhaps the simplest working example is given by three perturbative terms, such as
\begin{equation}\label{k_pert}
    k=\frac{a_1}{\bs+\overbar\bs}+\frac{a_2}{(\bs+\overbar\bs)^2}+\frac{a_3}{(\bs+\overbar\bs)^3}~.
\end{equation}
With this choice, the equations \eqref{Kpp_delta}, \eqref{Kp_delta}, and \eqref{Kppp_delta} can be solved perturbatively in $\delta$, which at the subleading order yields
\begin{align}\label{k_pert_sol}
\begin{aligned}
    a_1 &\simeq 6\langle\phi\rangle^2(1+4\sqrt{3}\langle\phi\rangle\,\delta)~,\\
    a_2 &\simeq -4\langle\phi\rangle^3(1+4\sqrt{3}\langle\phi\rangle\,\delta)~,\\
    a_3 &\simeq \langle\phi\rangle^4(1+4\sqrt{3}\langle\phi\rangle\,\delta)~,
\end{aligned}
\end{align}
where it can be seen that none of the three parameters can be set to zero (at least when $\langle\phi\rangle\delta$ is small), since $\langle\phi\rangle=0$ corresponds to a pole in the K\"ahler metric. It is also possible to add a constant term to \eqref{k_pert}, but it cannot be used to replace one of the $a_i$ parameters, and will only change the solution quantitatively.

By using \eqref{k_pert_sol}, the fourth derivative of $K$ at the vacuum becomes
\begin{equation}
        \langle K^{(4)}\rangle=-\frac{6}{\langle\phi\rangle^4}-\frac{6\sqrt{3}}{\langle\phi\rangle^3}\delta+\co(\delta^2)~.
\end{equation}
When $\delta$ is small enough for a given $\langle\phi\rangle$, the $\langle K^{(4)}\rangle\Omega^3$-term of the expansion \eqref{Kp_series} will dominate over the lower-derivative terms, and the cubic constraint $\Omega^3=(\bs+\overbar\bs-\langle\phi\rangle)^3=0$ becomes a good low-energy approximation of the dilaton equation of motion \eqref{dilaton_super_EOM}, as can be explicitly seen from the component EOM for the dilaton, which coincides with \eqref{global_R_EOM_limit} when sending $\delta\rightarrow 0$.

The canonical dilaton mass, gravitino mass, and the F-term VEV for the vacuum solution \eqref{k_pert_sol} read
\begin{align}
    m_\phi/\mu &=\frac{3}{\langle\phi\rangle^{5/2}\delta^2}-\frac{3\sqrt{3}}{\langle\phi\rangle^{3/2}\delta}+\co(\delta^0)~,\label{m_dilaton}\\
    \langle m_{3/2}\rangle/\mu &=\frac{1}{2\sqrt{\langle\phi\rangle}}-\frac{3\sqrt{3\langle\phi\rangle}}{4}\delta+\co(\delta^2)~,\label{m_gravitino}\\
    \langle F\rangle/\mu &=-\frac{\sqrt{3}}{2\sqrt{\langle\phi\rangle}\delta}+\frac{9\sqrt{\langle\phi\rangle}}{4}+\co(\delta)~.\label{F_dilaton}
\end{align}
Let us also expand the K\"ahler potential around the vacuum $\langle S+\overbar S\rangle=\langle\phi\rangle$. We first redefine the dilaton-axion superfield as
\begin{equation}\label{S_dilaton_redef}
    \bs=\frac{\tilde\bs}{\delta}+\frac{\langle\phi\rangle}{2}~,
\end{equation}
where we shift it by $\langle\phi\rangle/2$, so that the new dilaton $\tilde S+\overbar{\tilde S}$ vanishes at the vacuum, while the factor of $1/\delta$ leads to the canonically normalized K\"ahler potential,
\begin{align}\label{K_dilaton_constrained}
\begin{aligned}
    K &=\langle K\rangle+\langle K'\rangle(\bs+\overbar\bs-\langle\phi\rangle)+\tfrac{1}{2}\langle K''\rangle(\bs+\overbar\bs-\langle\phi\rangle)^2\\
    &=\langle K\rangle+\sqrt{3}(\tilde\bs+\overbar{\tilde\bs})+\tfrac{1}{2}(\tilde\bs+\overbar{\tilde\bs})^2~,
\end{aligned}
\end{align}
where we used $\langle K'\rangle=\sqrt{3}\delta$ and $\langle K''\rangle=\delta^2$. The higher-order terms vanish thanks to the effective cubic constraint derived from the EOM \eqref{dilaton_super_EOM}, which we can write as 
\begin{equation}
    (\tilde\bs+\overbar{\tilde\bs})^3=0~,
\end{equation}
Its solution eliminates the dilaton $\tilde\phi$ (where the tilde denotes the components of the redefined dilaton-axion superfield $\tilde\bs$) in terms of $\tilde\chi$, $\tilde\varphi$, and $\tilde F$ according to Eq. \eqref{cubic_constr_comp_SG}. 

The constant term of the K\"ahler potential \eqref{K_dilaton_constrained} (second line), which is equal to $\langle K\rangle=-\log{(4\langle\phi\rangle)}+\co(\delta)$ in our example \eqref{k_pert} and \eqref{k_pert_sol}, can be absorbed by redefining $\mu\rightarrow e^{-\langle K\rangle/2}\mu$. We then recover the minimal supergravity model \eqref{KW_constrained_SUGRA} for the cubic constraint, with $\xi=\sqrt{3}$ which corresponds to Minkowski vacuum. Note that the auxiliary field of the redefined dilaton-axion superfield $\tilde\bs$, is finite at the vacuum as $\delta$ is sent to zero,
\begin{equation}\label{F_tilde_value}
    \langle\tilde F\rangle/\mu=-\sqrt{3}+\co(\delta)~,
\end{equation}
which is obtained from \eqref{F_dilaton} by using $F=\tilde F/\delta$ and $\mu\rightarrow e^{-\langle K\rangle/2}\mu$. The expression \eqref{F_tilde_value} matches the auxiliary EOM \eqref{aux_EOM} of the minimal constrained model, if we set $\xi=\sqrt{3}$ (Minkowski vacuum) in \eqref{aux_EOM}.

Although in this model we consider a constant superpotential, the analysis can be extended to the case of an exponential (instanton-like) superpotential, such as $W\propto e^{-bS}$ or $W\propto \mu+e^{-bS}$. In the former case we have an exact R-symmetry, while in the latter case the R-symmetry is explicitly broken. This breaking must be sufficiently mild if we want to preserve the axion (i.e. $m_{\rm axion}\ll m_{\rm dilaton}$) and derive the cubic constraint.

\begin{figure}
\centering
  \includegraphics[width=.4\linewidth]{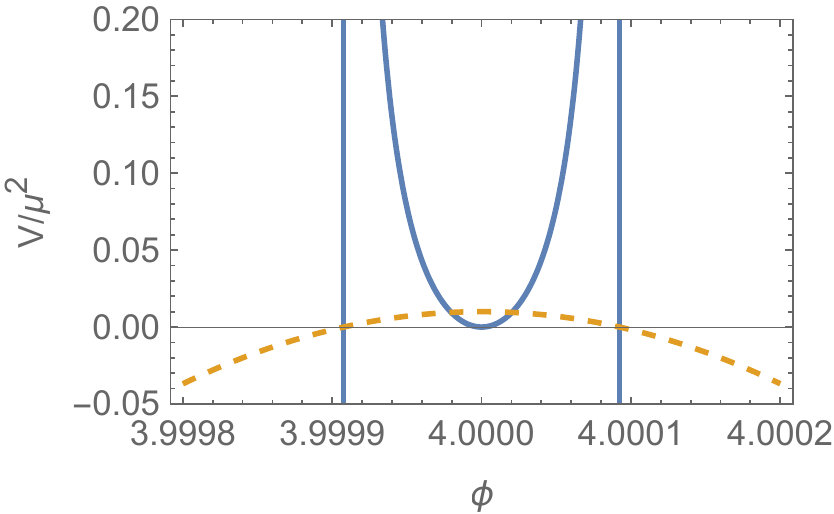}
\captionsetup{width=1\linewidth}
\caption{Dilaton scalar potential (solid curve) for the model \eqref{dilaton_KW} and \eqref{k_pert}. Dashed curve is the K\"ahler metric $K''$ multiplied by $10^8$ (to amplify its shape). Vertical lines are singularities of the scalar potential, where the K\"ahler metric vanishes.}
\label{Fig_V_dilaton}
\end{figure}

{\bf Explicit example.} Finally, let us confirm our perturbative analysis of the dilaton model numerically. If we take $\langle\phi\rangle=4$ and $\delta=10^{-5}$ as an example, the three parameters become,
\begin{equation}
a_1\approx 96.0266~,~~~a_2=-a_3\approx -256.071~.
\end{equation}
The resulting scalar potential is shown in Figure \ref{Fig_V_dilaton} as the solid blue curve. Stable Minkowski minimum can be seen, surrounded by two singularities coming from vanishing K\"ahler metric. The latter is shown as the dashed orange curve (amplified by $10^8$ to make its shape distinguishable), and inside the vertical lines $K''$ is positive, while outside - negative, which indicates unphysical field space. In this example the (canonical) dilaton mass is approximately equal to $0.94\times 10^{9}\mu$, and the gravitino mass to $\mu/4$, which can also be seen from Eqs. \eqref{m_dilaton} and \eqref{m_gravitino}, respectively.

Choosing different values of the dilaton VEV leads to different parameter sets $\{a_1,a_2,a_3\}$ for the correction term \eqref{k_pert}. For example for the smaller value of $\langle\phi\rangle=3$ we also have somewhat smaller parameter values
\begin{equation}
    a_1=~54(1+12\sqrt{3}\,\delta),~~~a_2=-108(1+12\sqrt{3}\,\delta)~,~~~a_3=81(1+12\sqrt{3}\,\delta)~.
\end{equation}
If we want the parameters to be of order $\co(1)$, however, the dilaton VEV may enter non-perturbative range $\langle\phi\rangle<1$ (for the K\"ahler corrections).

{\bf Other choices of $k$.} Aside from the ansatz \eqref{k_pert}, we tested several combinations of perturbative and non-perturbative corrections. We found that a simple combination
\begin{equation}
    k=\frac{a}{(\bs+\overbar\bs)^p}+\gamma e^{-\beta (\bs+\overbar\bs)^q}~,
\end{equation}
with positive $p$, $q$, and $\beta$, produces similar results. For example for $p=q=1$ and $\langle\phi\rangle=4$, we have
\begin{equation}
    a=48(1+23.09\,\delta)~,~~~\gamma=-53.56(1+23.09\,\delta)~,~~~\beta=3/4~.
\end{equation}
at the subleading order, by perturbatively solving \eqref{Kpp_delta}, \eqref{Kp_delta}, and \eqref{Kppp_delta}. On the other hand, for $p=1$ and $q=2$, we obtain substantially smaller parameter values (also for $\langle\phi\rangle=4$),
\begin{equation}
    a=2.94(1+18.48\,\delta)~,~~~\gamma=7.81(1+18.48\,\delta)~,~~~\beta=0.0172~.
\end{equation}
For both of these cases, with small enough $\delta$, the scalar potential (and K\"ahler metric) is similar in shape to that of Figure \ref{Fig_V_dilaton}, while the mass spectrum has similar $\delta$-expansion to Eqs. \eqref{m_dilaton}, \eqref{m_gravitino}, and \eqref{F_dilaton}, but with different numerical coefficients.

\subsection{Dilaton model with gauged $U(1)$}\label{Subsec_dilaton_gauge}

We can gauge the shift-symmetry of the dilaton-axion superfield $\bs$, and work in the superfield unitary gauge where we have a single massive vector superfield $\hat\bv$. This leads to the D-term-extended scalar potential,
\begin{equation}
    V=\mu^2e^J\left(\frac{{J'}^2}{J''}-3\right)+\frac{g^2}{2}{J'}^2~.
\end{equation}
As before, we introduce a small parameter $\delta$ as $\langle J''\rangle\equiv\delta^2$. Minkowski vacuum equations then yield
\begin{equation}\label{Jp_cond}
    \langle J'\rangle=\frac{\sqrt{3}\delta}{\sqrt{1+\tfrac{1}{2}\hat g^2\delta^2}}~,
\end{equation}
and
\begin{equation}\label{Jppp_cond}
    \langle J'''\rangle=\frac{2+\tfrac{1}{2}\hat g^2\delta^2+\tfrac{1}{2}\hat g^4\delta^4}{\sqrt{3}\sqrt{1+\tfrac{1}{2}\hat g^2\delta^2}}\delta^3~,
\end{equation}
where we denote $\hat g\equiv ge^{-\langle J\rangle/2}/\mu$. If $\hat g$ is as large as $1/\delta$, we cannot expand the above expressions in $\delta$.

The mass spectrum of this model, under the conditions \eqref{Jp_cond} and \eqref{Jppp_cond}, is given by
\begin{gather}
\begin{gathered}
    m^2_\phi=-6\mu^2e^{\langle J\rangle}\frac{(\langle J^{(4)}\rangle+\ldots)}{\delta^4(1+\tfrac{1}{2}\hat g^2\delta^2)}+2g^2\delta^2\big(1+\frac{1}{\delta^4}\langle J'\rangle\langle J'''\rangle\big)~,\\
    m_{\hat A}=\sqrt{2}g\delta~,~~~m_{1/2}=\frac{g^2\delta^2}{\mu}e^{-\langle J\rangle/2}~,~~~m_{3/2}=\mu e^{\langle J\rangle/2}~,\\
    \langle f\rangle=-\frac{\sqrt{3}\mu e^{\langle J\rangle/2}}{\delta\sqrt{1+\tfrac{1}{2}\hat g^2\delta^2}}~,~~~\langle D\rangle=-\frac{\sqrt{3}g\delta}{\sqrt{1+\tfrac{1}{2}\hat g^2\delta^2}}~,
\end{gathered}
\end{gather}
where the ellipsis in the $m_\phi$ expression stands for other terms depending on the lower derivatives of $J$ which provide subleading contributions to the $\langle J^{(4)}\rangle$-term (subleading in pure $\delta$, not counting the product $\hat g\delta$ which may not be small compared to $\langle J^{(4)}\rangle$ as explained below). Since $g$ here always enters as the product $g\delta$ (this is the case for the whole Lagrangian), if we take $\delta\rightarrow 0$ while $g$ is fixed, we obtain the decoupling limit of the dilaton multiplet and the gauge multiplet. Therefore we assume (formally) that $g$ behaves like $1/\delta$ as $\delta\rightarrow 0$.~\footnote{Taking the limit $\delta\rightarrow 0$ means that we ignore small corrections and keep only the leading-order expressions in $\delta$. However, quantities (such as $m_{\hat A}$) proportional to a positive power of $\delta$ may not be negligible in this limit compared to a relevant mass scale of the model. For example, $m_{\hat A}\sim g\delta$ can be comparable in magnitude to the parameter $\mu$ (in Planck units), even if $g$ is small, but formally, when sending $\delta$ to zero, we can say that $g\sim 1/\delta$ in order to keep $g\delta$ fixed.} Thus, the leading contribution in the dilaton mass (squared) is given by the $\langle J^{(4)}\rangle$-term, which is proportional to $\delta^{-4}$.

In order to reproduce the cubic constraint, we take a look at the superfield EOM. Since we established that the gauge coupling must behave non-trivially as we send $\delta$ to zero, we must restore it before analyzing the superfield EOM. To do this, we decompose $\hat\bv$ back into the massless gauge superfield $\bv$ and the chiral dilaton-axion superfield $\bs$. We include the gauge coupling by considering the K\"ahler potential as a function of $\bs+\overbar\bs+2g\bv$. The $\bv$-EOM in this case reads
\begin{equation}
    2ge^{-K/3}K'=\cd^\alpha\boldsymbol\cw_\alpha~,
\end{equation}
where $'$ stands for the derivative w.r.t. $\bs+\overbar\bs+2g\bv$. We also have the superfield EOM for $\bs$,
\begin{equation}
    (\overbar\cd^2-8\boldsymbol\car)(e^{-K/3}K')=0~.\label{Dilaton_gauged_super_EOM}
\end{equation}
Here the constraint can be readily seen if we use the expansion \eqref{Kp_series} for $K'$, and take the limit $\delta\rightarrow 0$ (so that $\langle K'\rangle$, $\langle K''\rangle$, and $\langle K'''\rangle$ vanish). From \eqref{Dilaton_gauged_super_EOM} we get
\begin{equation}
    (\bs+\overbar\bs+2g\bv-\langle \phi\rangle)^3=0~,
\end{equation}
where $\phi=S+\overbar S$. Since $\langle K''\rangle=\delta^2$, we want to canonically normalize the effective K\"ahler potential for $\bs$. As in the previous subsection, we introduce the canonical (and shifted) superfield $\tilde\bs$ as
\begin{equation}\label{S_dilaton_redef2}
    \bs=\frac{\tilde\bs}{\delta}+\frac{\langle\phi\rangle}{2}~,
\end{equation}
and the constraint becomes
\begin{equation}
    (\tilde\bs+\overbar{\tilde\bs}+2\tilde g\bv)^3=0~,
\end{equation}
where we introduced the redefined gauge coupling $\tilde g\equiv g\delta$, which stays finite when $\delta\rightarrow 0$. We can now reassemble the combination $\tilde\bs+\overbar{\tilde\bs}+2\tilde g\bv$ as a cubic nilpotent massive vector superfield. Let us summarize the mass spectrum when using the redefined dilaton multiplet $\tilde\bs$ and the gauge coupling $\tilde g$, and rescaling $\mu\rightarrow\mu e^{-\langle J\rangle/2}$,
\begin{gather}\label{Dilaton_gauged_masses_final}
\begin{gathered}
    m_{\hat A}=\sqrt{2}\tilde g~,~~~m_{1/2}=\tilde g^2/\mu~,~~~m_{3/2}=\mu~,\\
    \langle \tilde F\rangle=-\frac{\sqrt{6}\mu^2}{\sqrt{2\mu^2+\tilde g^2}}~,~~~\langle D\rangle=-\frac{\sqrt{6}\mu\tilde g}{\sqrt{2\mu^2+\tilde g^2}}~,
\end{gathered}
\end{gather}
while $m_\phi\rightarrow\infty$ when $\delta\rightarrow 0$, and the dilaton decouples. These results are consistent with the minimal model for the nilpotent massive vector superfield discussed in Subsection \ref{Subsec_nilpotent_V_Lagr}. They can also be compared to the UV model with gauged R-symmetry shown in Subsection \ref{subsec_local_R_UV}. Unlike the masses and SUSY breaking parameters \eqref{UV_gauge_R_masses} of the latter, the expressions of \eqref{Dilaton_gauged_masses_final} are exact, i.e. there are no $\delta$-corrections, since we used the exact forms of $\langle J'\rangle$, $\langle J''\rangle$, and $\langle J'''\rangle$ from the Minkowski vacuum equations ($J'$ in the massive vector formulation corresponds exactly to $K'$).

Explicit examples of the K\"ahler potential in this case are direct generalizations of those given in the previous subsection. As an example, we take the perturbative ansatz for the K\"ahler correction terms,
\begin{equation}
    K=-\log\left(S+\overbar S+\frac{a_1}{S+\overbar S}+\frac{a_2}{(S+\overbar S)^2}+\frac{a_3}{(S+\overbar S)^3}\right)~,
\end{equation}
while the superpotential is $W=\mu$. Solving $\langle K''\rangle=\delta^2$, and the Minkowski vacuum conditions \eqref{Jp_cond} and \eqref{Jppp_cond} up to $\co(\delta)$ yields
\begin{align}
\begin{aligned}
    a_1 &\simeq 6\langle\phi\rangle^2\left(1+\frac{4\sqrt{6}\langle\phi\rangle\mu\delta}{\sqrt{2\mu^2+\tilde g^2}}\right)~,\\
    a_2 &\simeq -4\langle\phi\rangle^3\left(1+\frac{4\sqrt{6}\langle\phi\rangle\mu\delta}{\sqrt{2\mu^2+\tilde g^2}}\right)~,\\
    a_3 &\simeq \langle\phi\rangle^4\left(1+\frac{4\sqrt{6}\langle\phi\rangle\mu\delta}{\sqrt{2\mu^2+\tilde g^2}}\right)~,
\end{aligned}
\end{align}
which reduces to \eqref{k_pert_sol} when $\tilde g=0$.

\subsection{Comment on anomalies}

Let us briefly discuss possible anomalies in the UV models with gauged shift-symmetry. 
There can be string-theoretic or field-theoretic anomalies, both of which can be cancelled by a Green--Schwarz mechanism, where we introduce a non-trivial gauge kinetic function which appropriately shifts under the gauge transformation. In this work we considered two UV models with gauged $U(1)$ symmetry: one with gauged R-symmetry (Subsection \ref{subsec_local_R_UV}), and one with non-R-symmetry (Subsection \ref{Subsec_dilaton_gauge}). In the latter case, since the dilaton-axion superfield $\bs$ transforms by a shift under the gauge symmetry, both fermions (dilatino and gaugino) are neutral, and there are no field-theoretic anomalies. In the model with gauged R-symmetry however, even if the chiral superfield transforms by a shift under the gauge symmetry, the fermions are charged, which can be seen from the non-invariance of the superspace $\theta$-coordinate. Therefore a Green--Schwarz mechanism must be implemented in this case. On the other hand, string-theoretic anomalies often accompany the dilaton-axion multiplet in e.g. heterotic string models, also requiring a Green--Schwarz cancellation mechanism.

Let us consider the effect of the field-dependent gauge kinetic function on the superfield EOM. We consider a general model of a chiral superfield $\bs$ with a gauged shift-symmetry, in the presence of gauge kinetic function $h=h(\bs)$ and $h'\neq 0$. More specifically $h$ must be linear in $\bs$ for the anomaly cancellation. Assuming that the K\"ahler potential depends on $\bs+\overbar\bs+\bv$, we write the superfield EOM,
\begin{align}
    \bv:&~~~e^{-K/3}K'=\tfrac{1}{2}(h+\bar h)\cd^\alpha\boldsymbol\cw_\alpha+\tfrac{1}{2}(h'\cd^\alpha\bs\boldsymbol\cw_\alpha+{\rm h.c.})~,\label{anomalous_superfield_EOM_V}\\
    \bs:&~~~(\overbar\cd^2-8\boldsymbol\car)(e^{-K/3}K')=h'\boldsymbol\cw^\alpha\boldsymbol\cw_\alpha~.\label{anomalous_superfield_EOM_S}
\end{align}
We discussed two alternative approaches to obtain the cubic constraint by decoupling the real scalar $S+\overbar S$. The first approach relies on the regime where $\langle K^{(4)}\rangle\rightarrow\infty$ while the lower derivatives are finite or vanishing. In this case, the effective cubic constraint can be obtained from the above EOM, as usual, by expanding $K'$ around the VEV of $S$ in \eqref{anomalous_superfield_EOM_V} and \eqref{anomalous_superfield_EOM_S}. Then $\langle K^{(4)}\rangle$ multiplies $(\bs+\overbar\bs+\bv-\langle S+\overbar S\rangle)^3$, and the latter effectively vanishes when taking $\langle K^{(4)}\rangle$ to infinity, provided that $h$ and $h'$ remain finite in this regime. This approach is used in our local $U(1)_R$ model of Subsection \ref{subsec_local_R_UV}. As can be seen, the presence of $h'\neq 0$ does not affect the results here. 

Next, consider the second approach which relies on vanishing $\langle K'\rangle$, $\langle K''\rangle$, and $\langle K'''\rangle$, and finite $\langle K^{(4)}\rangle$. In this case, we encounter an obstacle when trying to derive the cubic constraint: upon expanding $K'$ in Eqs. \eqref{anomalous_superfield_EOM_V} and \eqref{anomalous_superfield_EOM_S}, and taking $\langle K'\rangle,\langle K''\rangle,\langle K'''\rangle$ to zero, in addition to the surviving $\langle K^{(4)}\rangle$-term, we have terms proportional to $h$ and $h'$, involving the superfield strength $\boldsymbol\cw_\alpha$. If these terms are non-vanishing, we cannot obtain the constraint. One possible way out in this case, is to consider quantum corrections to the tree-level EOM \eqref{anomalous_superfield_EOM_V} and \eqref{anomalous_superfield_EOM_S}. These are also needed if we want to make the EOM gauge-invariant (\eqref{anomalous_superfield_EOM_V} is not invariant w.r.t. the gauge transformation of $\bs$ due to the non-vanishing derivative of $h$). We leave the study of quantum-corrected EOM for future research.

\section{Conclusion}\label{sec_concl}

In Ref. \cite{Aldabergenov:2021obf} new cubic constraints were proposed for $\cn=1$ chiral superfields, which can be used to construct low-energy effective theories with spontaneously broken (exact or approximate) global abelian symmetry and non-linear global $\cn=1$ supersymmetry. The constraints generate effective axion-goldstino interactions by eliminating the saxion (in terms of the axion and goldstino). In this work we generalized the results of \cite{Aldabergenov:2021obf} to the case with local abelian symmetry, as well as to supergravity. In the former case we showed that in the presence of a vector multiplet gauging the abelian symmetry, the construction can be described (in the $U(1)$ unitary gauge) by a cubic nilpotent massive vector superfield ($\hat\bv^3=0$) whose real scalar component is eliminated by the constraint. The supergravity generalization of these constraints is straightforward, and it extends the global SUSY solutions by (old-minimal) supergravity multiplet components. We derived these explicitly in Section \ref{sec_cubic_SUGRA} for the cases with both global and local abelian symmetries. 

For each constrained superfield we constructed a minimal Lagrangian, and we found that the minimal (single-superfield) model for the cubic nilpotent massive vector superfield $\hat\bv$ is inconsistent in global SUSY, due to the vanishing auxiliary F-term of $\hat\bv$ in which case the solution to the constraint becomes singular. In supergravity however, this problem is resolved in the presence of a constant non-zero superpotential. 

Lastly, we discussed several examples of UV completions of these constrained superfields, including a supergravity toy model for the string dilaton with global or local $U(1)$ which shifts the axion. When taking an appropriate low-energy limit in these models, the saxion becomes infinitely heavy and decouples, while the axion is protected by the shift-symmetry. SUSY breaking scale, or more precisely the quantity $(F\overbar F-\tfrac{1}{4}\partial\varphi\partial\varphi)^{1/2}$ which includes the axion kinetic energy, provides a cutoff scale for the effective theory. When the abelian symmetry is gauged, we also obtain a massive spin-1/2 fermion, while the axion combines with the gauge field to form a massive vector. 

One interesting consequence of the cubic constraints and their (minimal) supergravity models, is that the abelian symmetry necessarily becomes an R-symmetry, even if in the microscopic model we start from non-R $U(1)$, like in the dilaton case. This can be seen from the expansion of the K\"ahler potential (assuming a constant superpotential),
\begin{equation}
    K=\langle K\rangle+\langle K'\rangle (\mathbf S+\overbar{\mathbf  S})+\tfrac{1}{2}\langle K''\rangle(\mathbf S+\overbar{\mathbf  S})^2~,~~~(\mathbf S+\overbar{\mathbf  S})^3=0~,
\end{equation}
where we assume vanishing VEV of the chiral scalar $S$. Since the auxiliary field $F\propto K'\neq 0$, the linear term in $K$ is always present when the constraint is imposed, and it is equivalent (up to a K\"ahler transformation) to having a charged superpotential $W\propto e^{\langle K'\rangle\mathbf  S}$. When the shift-symmetry is gauged, the constant $\langle K'\rangle$ provides an effective $U(1)_R$ Fayet--Iliopoulos term, which can be useful for phenomenological model building.

\section*{Acknowledgements}

This work was supported in part by the NSRF via the Program Management Unit for Human Resources and Institutional Development, Research and Innovation [grant number B01F650006 and B05F650021]. A.C. is also supported in part by Thailand Science research and Innovation Fund Chulalongkorn University (IND66230009).

\appendix

\section{Supergravity action}
We use curved superspace formalism of \cite{Wess:1992cp}. The superspace action for a chiral superfield $\bs$ coupled to a $U(1)$ vector superfield $\bv$ reads ($M_{\rm P}=1$)
\begin{equation}
    {\cal L}=\int d^2\Theta 2{\cal E}\left[\tfrac{3}{8}(\overbar{\cal D}^2-8\boldsymbol{\cal R})e^{-K/3}+W+\tfrac{1}{4}h\boldsymbol{\cal W}^\alpha\boldsymbol{\cal W}_\alpha\right]+{\rm h.c.}~,\label{App_L_super}
\end{equation}
where $\cal E$ is chiral density superfield, $\cal R$ is chiral curvature superfield, ${\cal D}_\alpha,\overbar{\cal D}_{\dot{\alpha}}$ are the superspace covariant derivatives with ${\cal D}^2\equiv{\cal D}^\alpha{\cal D}_\alpha$ and $\overbar{\cal D}^2\equiv\overbar{\cal D}_{\dot{\alpha}}\overbar{\cal D}^{\dot{\alpha}}$. Here $K=K({\bs},\overbar{\bs},\bv)$ is gauge-invariant superfield K\"ahler potential,  $W=W(\bs)$ is superpotential, $h=h(\bs)$ is gauge kinetic function, and $\boldsymbol{\cal W}_\alpha\equiv -\tfrac{1}{4}(\overbar{\cal D}^2-8\boldsymbol{\cal R}){\cal D}_\alpha{\bf V}$ is (chiral) superfield strength of $\bv$. The operator $(\overbar{\cal D}^2-8\boldsymbol{\cal R})$ acts as a chiral projector in curved superspace, such that $\overbar\cd_{\dot\alpha}(\overbar{\cal D}^2-8\boldsymbol{\cal R})=0$ when acting on a general superfield. 

We denote the Killing vector of the gauge symmetry in $S$-direction as $X^S$, and the Killing potential is
\begin{equation}\label{Killing_pot}
\mathscr{D}=iX^S\left(K_S+\frac{W_S}{W}\right)~,
\end{equation}
where the second term is the Fayet-Iliopoulos term of gauged $R$-symmetry, and vanishes if the $U(1)$ is not R-symmetry. For the gauge coupling we use the letter $g$ in our notation.

The expansion of $\boldsymbol\ce$ and $\boldsymbol\car$ in $\Theta$-variable is
\begin{align}
\begin{split}\label{E_expansion}
    2\boldsymbol\ce=&~e\left[1+i\Theta\sigma^m\bar\psi_m-\Theta^2(\overbar M+\bar\psi_m\overbar\sigma^{mn}\bar\psi_n)\right]~,
\end{split}\\
\begin{split}\label{R_expansion}
    -6\boldsymbol\car=&~M+\Theta(\sigma^m\overbar\sigma^n\psi_{mn}-i\sigma^m\bar\psi_m M+i\psi_m b^m)\\
    &+\Theta^2\left[\frac{1}{2}R+i\bar\psi^m\overbar\sigma^n\psi_{mn}+\frac{2}{3}M\overbar{M}+\frac{1}{3}b_m b^m-ie^m_a\cd_m b^a\right.\\
    &+\left.\frac{1}{2}\bar\psi_m\bar\psi^m M-\frac{1}{2}\psi_m\sigma^m\bar\psi_n b^n+\frac{1}{8}\varepsilon^{klmn}(\bar\psi_k\overbar\sigma_l\psi_{mn}+\psi_k\sigma_l\bar\psi_{mn})\right]~,
\end{split}
\end{align}
where $e\equiv{\rm det}\,e^a_m$, $e_m^a$ and $\psi_m$ are graviton (frame field) and gravitino, and $M$ (complex scalar) and $b_m$ (real vector) are old-minimal SUGRA auxiliary fields. The covariant derivative $\cd_m$ acts on local Lorentz indices,
\begin{align}
\begin{aligned}
    \cd_m V_a &=\partial_a+V^b\omega_{mba}~,\\
    \cd_m U^\alpha &=\partial_mU^\alpha+U^\beta{\omega_{m\beta}}^\alpha~,\\
    \cd_m \overbar U_{\dot\alpha} &=\partial_m\overbar U_{\dot\alpha}+\overbar U_{\dot\beta}{{\omega_m}^{\dot\beta}}_{\dot\alpha}~.
\end{aligned}
\end{align}
$\{a,b,c,\ldots\}$ are Lorentz spacetime indices, $\{\alpha,\beta,\gamma,\ldots\}$ are Lorentz spinor indices. The connection components $\omega_{mnk}$ are given by
\begin{align}
\begin{aligned}\label{spin_con}
    \omega_{mnk}&=\tfrac{1}{2}[e_{ma}\partial_{(n}e^a_{k)}-e_{na}\partial_{(k}e^a_{m)}-e_{ka}\partial_{(m}e^a_{n)}]\\
    &+\tfrac{i}{4}[e_{ma}\psi_{(k}\sigma^a\bar\psi_{n)}-e_{na}\psi_{(m}\sigma^a\bar\psi_{k)}-e_{ka}\psi_{(n}\sigma^a\bar\psi_{m)}]~,
\end{aligned}
\end{align}
and
\begin{equation}
    {\omega_{m\beta}}^\alpha=-\tfrac{1}{2}\omega_{mnk}{{\sigma^{nk}}_\beta}^\alpha~,~~~{{\omega_m}^{\dot\beta}}_{\dot\alpha}=-\tfrac{1}{2}\omega_{mnk}{\overbar\sigma^{nk\dot\beta}}_{\dot\alpha}~,
\end{equation}
where in \eqref{spin_con} we use $A_{(m}B_{n)}\equiv A_mB_n-A_nB_m$. Note that the connection includes gravitino contributions responsible for non-zero torsion. Covariant divergence acting on spacetime vector index is equal to
\begin{equation}
    \nabla_mV^n=\partial_mV^n+\Gamma^n_{mk}V^k=e^n_a\cd_mV^a~,
\end{equation}
where the affine connection can be expressed as $\Gamma^n_{mk}=\Gamma^n_{mk}(e)+\Gamma^n_{mk}(\psi)$, with Christoffel symbols $\Gamma^n_{mk}(e)$ and the gravitino-dependent part $\Gamma^n_{mk}(\psi)={\omega_{mk}}^n(\psi)$ (the latter is given by the second line of \eqref{spin_con}). In \eqref{E_expansion} and \eqref{R_expansion} we also used the notation
\begin{equation}
    \psi^\alpha_{mn}\equiv \cd_m\psi^\alpha_n-\cd_n\psi^\alpha_m~,
\end{equation}
for the gravitino field strength, and $\sigma^{mn}\equiv\tfrac{1}{4}\sigma^{(m}\bar\sigma^{n)}$.

We define the (supercovariant) scalar curvature $R$ in terms of the spin connection $\omega$,
\begin{equation}
    R=e^m_ae^n_b[\partial_{(n}{\omega_{m)}}^{ab}+{\omega_{(m}^{ac}\omega_{n)c}}^b]~.
\end{equation}

The component Lagrangian in Jordan frame contains the curvature term $ee^{-K/3}R/2$. The Einstein frame can be obtained by using simultaneous redefinitions
\begin{equation}
    e_m^a\rightarrow e^{K/6}e_m^a~,~~~\psi_m\rightarrow e^{K/12}\psi_m~.
\end{equation}
The curvature term transforms as
\begin{equation}
    \tfrac{1}{2}ee^{-K/3}R\rightarrow\tfrac{1}{2}eR-\tfrac{1}{12}e\partial_mK\partial^mK+\tfrac{i}{6}e(\psi_n\sigma^m\bar\psi_m-\psi_m\sigma^m\bar\psi_n)\partial_nK-\tfrac{1}{2}\partial_m(e\partial^mK)~.
\end{equation}
When present, we also rescale the chiral fermion $\chi\rightarrow e^{-K/12}\chi$, gaugino $\lambda\rightarrow e^{-K/4}\lambda$, and the auxiliary fields $F\rightarrow e^{-K/6}F$ and $D\rightarrow e^{-K/3}D$ for appropriate normalization.

In order to eliminate kinetic terms with $\psi_m$-$\chi$ mixing, the gravitino is shifted as
\begin{equation}\label{psi_shift}
    \psi^\alpha_m\rightarrow\psi^\alpha_m-\tfrac{\sqrt{2}}{6}iK'\bar\chi_{\dot\alpha}\overbar\sigma_m^{\dot\alpha\alpha}~.
\end{equation}
Here we assume that $K$ is a function of $S+\overbar S$, and so $K'=K_S=K_{\bar S}$. \eqref{psi_shift} leads to the transformation of the supercovariant scalar curvature,
\begin{align}
    \tfrac{1}{2}eR& \rightarrow\tfrac{1}{2}eR+\tfrac{\sqrt{2}i}{12}eK'\big[(\chi\psi^n-\chi\sigma^{mn}\psi_m)(\psi_k\sigma^k\bar\psi_n-\psi_n\sigma^k\bar\psi_k)+\chi\sigma^{mn}\psi^k(\psi_m\sigma_k\bar\psi_n)-{\rm h.c.}\big]\nonumber\\
    &+\tfrac{i}{72}e{K'}^2\epsilon^{mnkl}\psi_m\sigma_n\bar\psi_k\chi\sigma_l\bar\chi+\tfrac{1}{144}e{K'}^2(3\chi^2\psi_m\psi^m-6\chi\psi_m\bar\chi\bar\psi^m-8\chi\psi_m\bar\chi\overbar\sigma^{mn}\bar\psi_n+{\rm h.c.})\nonumber\\
    &+\tfrac{\sqrt{2}}{144}e{K'}^3(i\chi^2\psi_m\sigma^m\bar\chi+{\rm h.c.})-\tfrac{1}{216}e{K'}^4\chi^2\bar\chi^2~,
\end{align}
up to total derivatives.


\section{Chiral and vector superfield components}

The components of a chiral superfield $\bs$ are the same as in global SUSY
\begin{equation}
    \bs=S+\sqrt{2}\Theta\chi+\Theta^2F~. \label{S_expansion}
\end{equation}
Or equivalently we can define them by using $\cd_\alpha$ and $\overbar\cd_{\dot\alpha}$,
\begin{equation}\label{app_S_def}
	\bs|=S~,~~~\cd_\alpha\bs|=\sqrt{2}\chi_\alpha~,~~~\cd^2\bs|=-4F~.
\end{equation}
From these we can obtain the following mixed derivatives by using the supergravity algebra of $\cd_\alpha$ and $\overbar\cd_{\dot\alpha}$ \cite{Wess:1992cp},
\begin{align}
\begin{aligned}
	\cd_\alpha\overbar\cd_{\dot\alpha}\overbar\bs| &=-2i\sigma^m_{\alpha\dot\alpha}\hat D_m \overbar S~\\
	\cd^2\overbar\cd_{\dot\alpha}\overbar\bs| &=-\tfrac{4\sqrt{2}}{3}\overbar M\bar\chi_{\dot\alpha}~,\\
	\cd_\alpha\overbar\cd^2\overbar\bs| &=-2\sqrt{2}\sigma^m_{\alpha\dot\alpha}\left(2i\hat D_m+\tfrac{1}{3}b_m\right)\bar\chi^{\dot\alpha}~,\\
	\cd^2\overbar\cd^2\overbar\bs| &=16e_a^m\cd_m\hat D^a\overbar S+\tfrac{32}{3}(ib^m\hat D_m\overbar S+\overbar M\overbar F)-8\sqrt{2}\bar\psi^m\hat D_m\bar\chi\\
	&\hspace{2cm}+\tfrac{8\sqrt{2}}{3}\left(\bar\psi_{mn}\bar\sigma^{mn}\bar\chi+i\bar\psi_m\bar\chi b^m-\tfrac{i}{2}\bar\psi_m\bar\sigma^m\sigma^n\bar\chi b_n\right)~.
\end{aligned}
\end{align}
Here we use the following supercovariant derivatives,
\begin{align}
\begin{aligned}
	\hat D_m\overbar S &\equiv \partial_m\overbar S-\tfrac{1}{\sqrt{2}}\bar\psi_m\bar\chi~,\\
	\hat D_m\bar\chi_{\dot\alpha} &\equiv\cd_m\bar\chi_{\dot\alpha}+\tfrac{i}{\sqrt{2}}\psi_m^\alpha\sigma^n_{\alpha\dot\alpha}\hat D_n\overbar S-\tfrac{1}{\sqrt{2}}\bar\psi_{m\dot\alpha}\overbar F~.
\end{aligned}
\end{align}

The components of an abelian vector superfield $\bv$ can be defined as follows. The definitions of the lower components are motivated by the components of the real superfield $\bs+\overbar\bs$,
\begin{gather}
	\bv|=\alpha~,~~~\cd_\alpha\bv|=\sqrt{2}\eta_\alpha~,~~~\cd^2\bv|=-4f~,\\
	\cd_\alpha\overbar\cd_{\dot\alpha}\bv|=-\sigma^m_{\alpha\dot\alpha}(A_m+i\partial_m\alpha-\sqrt{2}i\bar\eta\bar\psi_m)~,\label{vec_comp_SG}
\end{gather}
where $\alpha$ is a real scalar, $\eta$ is a Weyl fermion, $A_m$ is an abelian gauge field, and $f$ is complex auxiliary scalar. The gaugino $\lambda$ and the $D$-field can be defined by using the superfield strength $\boldsymbol\cw_\alpha=-\tfrac{1}{4}(\overbar\cd^2-8\boldsymbol\car)\cd_\alpha\bv$,
\begin{gather}
\begin{gathered}
	\boldsymbol\cw_\alpha|=-i\lambda_\alpha~,~~~\cd_\alpha \boldsymbol\cw^\beta|=i{{\sigma^{mn}}_{\alpha}}^{\beta}\hat F_{mn}+\delta^{\beta}_{\alpha}D~,\\
 \cd^2\boldsymbol\cw_\alpha|=-4\sigma^m_{\alpha\dot\beta}\hat D_m\bar\lambda^{\dot\beta}+2i(\lambda_\alpha\overbar M-\sigma^m_{\alpha\dot\beta}\lambda^{\dot\beta}b_m)~,
 \end{gathered}
\end{gather}
where $\hat F_{mn}$ is supercovariant field strength of $A_m$,
\begin{equation}
    \hat F_{mn}=F_{mn}+\tfrac{i}{2}(\lambda\sigma_n\bar\psi_m-\lambda\sigma_m\bar\psi_n-\psi_m\sigma_n\bar\lambda+\psi_n\sigma_m\bar\lambda)~,
\end{equation}
and $\hat D_m\bar\lambda$ is defined as
\begin{equation}
    \hat D_m\bar\lambda^{\dot\alpha}\equiv \cd_m\bar\lambda^{\dot\alpha}+\tfrac{i}{2}\bar\psi_m^{\dot\alpha}D-\tfrac{1}{2}{\overbar\sigma^{nk\,\dot\alpha}}_{\dot\beta}\bar\psi^{\dot\beta}_m\hat F_{nk}~.
\end{equation}
$F_{mn}$ is the regular field strength, $F_{mn}=\partial_mA_n-\partial_nA_m$.

We can then derive three- and four-derivative components of $\bv$,
\begin{align}
\begin{aligned}
	\cd^2\overbar\cd_{\dot\alpha}\bv| &=-4i\bar\lambda_{\dot\alpha}-\tfrac{4\sqrt{2}}{3}\overbar M\bar\eta_{\dot\alpha}~,\\
	\cd_\alpha\overbar\cd^2\bv| &=4i\lambda_\alpha-2\sigma^m_{\alpha\dot\alpha}\Big[2\sqrt{2}i\cd_m\bar\eta^{\dot\alpha}+\tfrac{\sqrt{2}}{3}\bar\eta^{\dot\alpha}b_m-2i\bar\psi_m^{\dot\alpha}\bar f\\
	&\hspace{4.2cm}+\overbar\sigma^{n\dot\alpha\beta}\psi_{m\beta}(\partial_n\alpha-iA_n-\sqrt{2}\bar\eta\bar\psi_n)\Big]~,\\
	\cd^2\overbar\cd^2\bv| &=8D+8e^m_a\cd_m(\partial^a\alpha-iA^a)+\tfrac{32}{3}\overbar M\bar f+\tfrac{16}{3}b^m(A_m+i\partial_m\alpha)-8\sqrt{2}\bar\psi^m\cd_m\bar\eta\\
	&-8\sqrt{2}e^m_a\cd_m(\bar\eta\bar\psi^a)+4\lambda\sigma^m\bar\psi_m+4\bar\lambda\overbar\sigma^m\psi_m-\tfrac{4\sqrt{2}}{3}i\bar\eta\bar\psi^mb_m-\tfrac{8\sqrt{2}}{3}\bar\eta\overbar\sigma^{mn}\bar\psi_{mn}\\
	&+\tfrac{8\sqrt{2}}{3}i\bar\eta\overbar\sigma^{mn}\bar\psi_mb_n+8\bar\psi^m\bar\psi_m\bar f-4i\psi_n\sigma^m\bar\psi^n(\partial_m\alpha-iA_m-\sqrt{2}\bar\eta\bar\psi_m)~.
\end{aligned}
\end{align}

Under the gauge symmetry, $\bv$ transforms as
\begin{equation}\label{V_transform}
	\bv\rightarrow\bv+\mathbf\Sigma+\overbar{\mathbf\Sigma}~,
\end{equation}
where $\mathbf\Sigma$ is a chiral superfield whose components we denote as
\begin{equation}
    \mathbf\Sigma=\rho+\sqrt{2}\Theta\kappa+\Theta^2\cf~.
\end{equation}
Extracting different components of \eqref{V_transform} we can infer the gauge transformations of all the component fields of $\bv$,
\begin{equation}
    \alpha\rightarrow\alpha+\rho+\bar\rho~,~~~\eta\rightarrow\eta+\kappa~,~~~f\rightarrow f+\cf~,~~~A_m\rightarrow A_m-i\partial_m(\rho-\bar\rho)~,
\end{equation}
while $\lambda$ and $D$ are invariant. In Wess--Zumino gauge we fix $\rho+\bar\rho$, $\kappa$, and $\cf$ such that $\alpha$, $\eta$, and $f$ vanish. This leaves the ordinary gauge transformation of $A_m$ by the derivative of $i(\rho-\bar\rho)$. It is worth noting that Wess--Bagger \cite{Wess:1992cp} uses a different definition of the component $\cd_\alpha\overbar\cd_{\dot\alpha}\bv|$ compared to Eq. \eqref{vec_comp_SG}. Namely, in their definition the gravitino-dependent term is absent, which leads to the ``unconventional" gauge transformation of $A_m$,
\begin{equation}\label{A_transform_kappa}
    A_m\rightarrow A_m-i\partial_m(\rho-\bar\rho)-\tfrac{i}{\sqrt{2}}(\kappa\psi_m-\bar\kappa\bar\psi_m)~,
\end{equation}
although this last term vanishes in the Wess-Zumino gauge. The role of the last term of \eqref{vec_comp_SG} is precisely to cancel the fermionic terms of \eqref{A_transform_kappa}, leaving only the conventional gauge transformation of $A_m$ even without Wess--Zumino gauge fixing.

In the case of a massive vector superfield, all of its components are physical and cannot be gauged away, since the gauge symmetry is already fixed by the choice $\Sigma=\bs$, where $\bs$ is the St\"uckelberg chiral superfield. In the main text we use $\hat\bv$ and $\hat A_m$ when describing massive vector superfield and its vector component, respectively.

\paragraph{A comment on general structure of cubic constraints.}
Whether supersymmetry is global or local, we are equipped with supersymmetric derivatives, which in this paragraph we denote by $\cd_\al,\overbar\cd_\dal$ for both cases. In the main text, constraint equations in component fields are obtained by acting on superfield constraints with $\cd^2\overbar\cd^2$ and setting the fermionic coordinates to zero. More concretely, for a superfield constraint $f(\boldsymbol\Phi^i)$ which is cubic in general superfields $\boldsymbol\Phi^i$,\footnote{We suppose that the leading component of each $\boldsymbol\Phi_i$ is bosonic.} the constraint equation in components $\cd^2\bar\cd^2f|$ reads,
\begin{align}\label{general-DDbDbD}
    &\cd^2\overbar\cd^2\boldsymbol\Phi^if_i|
    +\left(\overbar\cd^2\boldsymbol\Phi^i\cd^2\boldsymbol\Phi^j+2\cd_\al\overbar\cd_\dal\boldsymbol\Phi^i\cd^\al\overbar\cd^\dal\boldsymbol\Phi^j+2\cd^\al\overbar\cd^2\boldsymbol\Phi^i\cd_\al\boldsymbol\Phi^j+2\cd^2\overbar\cd_\dal\boldsymbol\Phi^i\overbar\cd^\dal\boldsymbol\Phi^j\right)f_{ij}| \nonumber\\
    &\quad=\left(4\cd_\al\overbar\cd_\dal\boldsymbol\Phi^i\overbar\cd^\dal\boldsymbol\Phi^j\cd^\al\boldsymbol\Phi^k-\overbar\cd^2\boldsymbol\Phi^i\cd^\al\boldsymbol\Phi^j\cd_\al\boldsymbol\Phi^k-\cd^2\boldsymbol\Phi^i\overbar\cd_\dal\boldsymbol\Phi^j\overbar\cd^\dal\boldsymbol\Phi^k\right)f_{kji}|,
\end{align}
where each suffix $i$ on $f$ means the differentiation with respect to $\boldsymbol\Phi^i$.
One can see that the left hand side of each of the constraints \eqref{cubic_constr_comp}, \eqref{cubic_constr_V}, \eqref{cubic_constr_comp_SG}, \eqref{nil_vec_comp_SG} comes from the first line of \eqref{general-DDbDbD}, while the right hand side from its second line. In particular, for these constraints, the terms with two fermionic fields, which come from $f_{ij}|$ and $f_{ijk}|$ parts in \eqref{general-DDbDbD}, enable us to obtain the solutions to the constraint equations in closed forms.

\providecommand{\href}[2]{#2}\begingroup\raggedright\endgroup


\begin{thebibliography}{10}

\bibitem{Rocek:1978nb}
M.~Rocek, ``{Linearizing the Volkov-Akulov Model},''
  \href{http://dx.doi.org/10.1103/PhysRevLett.41.451}{{\em Phys. Rev. Lett.}
  {\bfseries 41} (1978) 451--453}.

\bibitem{Ivanov:1978mx}
E.~A. Ivanov and A.~A. Kapustnikov, ``{General Relationship Between Linear and
  Nonlinear Realizations of Supersymmetry},''
  \href{http://dx.doi.org/10.1088/0305-4470/11/12/005}{{\em J. Phys. A}
  {\bfseries 11} (1978) 2375--2384}.

\bibitem{Lindstrom:1979kq}
U.~Lindstrom and M.~Rocek, ``{Constrained local superfields},''
  \href{http://dx.doi.org/10.1103/PhysRevD.19.2300}{{\em Phys. Rev. D}
  {\bfseries 19} (1979) 2300--2303}.

\bibitem{Ivanov:1982bpa}
E.~A. Ivanov and A.~A. Kapustnikov, ``{The nonlinear realization structure of
  models with spontaneously broken supersymmetry},''
  \href{http://dx.doi.org/10.1088/0305-4616/8/2/004}{{\em J. Phys. G}
  {\bfseries 8} (1982) 167--191}.

\bibitem{Samuel:1982uh}
S.~Samuel and J.~Wess, ``{A Superfield Formulation of the Nonlinear Realization
  of Supersymmetry and Its Coupling to Supergravity},''
  \href{http://dx.doi.org/10.1016/0550-3213(83)90622-3}{{\em Nucl. Phys. B}
  {\bfseries 221} (1983) 153--177}.

\bibitem{Casalbuoni:1988xh}
R.~Casalbuoni, S.~De~Curtis, D.~Dominici, F.~Feruglio, and R.~Gatto,
  ``{Nonlinear Realization of Supersymmetry Algebra From Supersymmetric
  Constraint},'' \href{http://dx.doi.org/10.1016/0370-2693(89)90788-0}{{\em
  Phys. Lett. B} {\bfseries 220} (1989) 569--575}.

\bibitem{Komargodski:2009rz}
Z.~Komargodski and N.~Seiberg, ``{From Linear SUSY to Constrained
  Superfields},'' \href{http://dx.doi.org/10.1088/1126-6708/2009/09/066}{{\em
  JHEP} {\bfseries 09} (2009) 066},
  \href{http://arxiv.org/abs/0907.2441}{{\ttfamily arXiv:0907.2441 [hep-th]}}.

\bibitem{Wess:1992cp}
J.~Wess and J.~Bagger, {\em {Supersymmetry and supergravity}}.
\newblock Princeton University Press, Princeton, NJ, USA, 1992.

\bibitem{Kuzenko:2010ef}
S.~M. Kuzenko and S.~J. Tyler, ``{Relating the Komargodski-Seiberg and
  Akulov-Volkov actions: Exact nonlinear field redefinition},''
  \href{http://dx.doi.org/10.1016/j.physletb.2011.03.020}{{\em Phys. Lett. B}
  {\bfseries 698} (2011) 319--322},
  \href{http://arxiv.org/abs/1009.3298}{{\ttfamily arXiv:1009.3298 [hep-th]}}.

\bibitem{Volkov:1973ix}
D.~V. Volkov and V.~P. Akulov, ``{Is the Neutrino a Goldstone Particle?},''
  \href{http://dx.doi.org/10.1016/0370-2693(73)90490-5}{{\em Phys. Lett. B}
  {\bfseries 46} (1973) 109--110}.

\bibitem{Ferrara:2015tyn}
S.~Ferrara, R.~Kallosh, and J.~Thaler, ``{Cosmology with orthogonal nilpotent
  superfields},'' \href{http://dx.doi.org/10.1103/PhysRevD.93.043516}{{\em
  Phys. Rev. D} {\bfseries 93} no.~4, (2016) 043516},
  \href{http://arxiv.org/abs/1512.00545}{{\ttfamily arXiv:1512.00545
  [hep-th]}}.

\bibitem{Kahn:2015mla}
Y.~Kahn, D.~A. Roberts, and J.~Thaler, ``{The goldstone and goldstino of
  supersymmetric inflation},''
  \href{http://dx.doi.org/10.1007/JHEP10(2015)001}{{\em JHEP} {\bfseries 10}
  (2015) 001}, \href{http://arxiv.org/abs/1504.05958}{{\ttfamily
  arXiv:1504.05958 [hep-th]}}.

\bibitem{DallAgata:2016syy}
G.~Dall'Agata, E.~Dudas, and F.~Farakos, ``{On the origin of constrained
  superfields},'' \href{http://dx.doi.org/10.1007/JHEP05(2016)041}{{\em JHEP}
  {\bfseries 05} (2016) 041}, \href{http://arxiv.org/abs/1603.03416}{{\ttfamily
  arXiv:1603.03416 [hep-th]}}.

\bibitem{Cribiori:2017ngp}
N.~Cribiori, G.~Dall'Agata, and F.~Farakos, ``{From Linear to Non-linear SUSY
  and Back Again},'' \href{http://dx.doi.org/10.1007/JHEP08(2017)117}{{\em
  JHEP} {\bfseries 08} (2017) 117},
  \href{http://arxiv.org/abs/1704.07387}{{\ttfamily arXiv:1704.07387
  [hep-th]}}.

\bibitem{Antoniadis:2014oya}
I.~Antoniadis, E.~Dudas, S.~Ferrara, and A.~Sagnotti, ``{The
  Volkov\textendash{}Akulov\textendash{}Starobinsky supergravity},''
  \href{http://dx.doi.org/10.1016/j.physletb.2014.04.015}{{\em Phys. Lett. B}
  {\bfseries 733} (2014) 32--35},
  \href{http://arxiv.org/abs/1403.3269}{{\ttfamily arXiv:1403.3269 [hep-th]}}.

\bibitem{Bergshoeff:2015tra}
E.~A. Bergshoeff, D.~Z. Freedman, R.~Kallosh, and A.~Van~Proeyen, ``{Pure de
  Sitter Supergravity},''
  \href{http://dx.doi.org/10.1103/PhysRevD.93.069901}{{\em Phys. Rev. D}
  {\bfseries 92} no.~8, (2015) 085040},
  \href{http://arxiv.org/abs/1507.08264}{{\ttfamily arXiv:1507.08264
  [hep-th]}}. [Erratum: Phys.Rev.D 93, 069901 (2016)].

\bibitem{Hasegawa:2015bza}
F.~Hasegawa and Y.~Yamada, ``{Component action of nilpotent multiplet coupled
  to matter in 4 dimensional $ \mathcal{N}=1 $ supergravity},''
  \href{http://dx.doi.org/10.1007/JHEP10(2015)106}{{\em JHEP} {\bfseries 10}
  (2015) 106}, \href{http://arxiv.org/abs/1507.08619}{{\ttfamily
  arXiv:1507.08619 [hep-th]}}.

\bibitem{Kuzenko:2015yxa}
S.~M. Kuzenko, ``{Complex linear Goldstino superfield and supergravity},''
  \href{http://dx.doi.org/10.1007/JHEP10(2015)006}{{\em JHEP} {\bfseries 10}
  (2015) 006}, \href{http://arxiv.org/abs/1508.03190}{{\ttfamily
  arXiv:1508.03190 [hep-th]}}.

\bibitem{Kallosh:2015tea}
R.~Kallosh and T.~Wrase, ``{De Sitter Supergravity Model Building},''
  \href{http://dx.doi.org/10.1103/PhysRevD.92.105010}{{\em Phys. Rev. D}
  {\bfseries 92} no.~10, (2015) 105010},
  \href{http://arxiv.org/abs/1509.02137}{{\ttfamily arXiv:1509.02137
  [hep-th]}}.

\bibitem{Ferrara:2015gta}
S.~Ferrara, M.~Porrati, and A.~Sagnotti, ``{Scale invariant
  Volkov\textendash{}Akulov supergravity},''
  \href{http://dx.doi.org/10.1016/j.physletb.2015.08.066}{{\em Phys. Lett. B}
  {\bfseries 749} (2015) 589--591},
  \href{http://arxiv.org/abs/1508.02939}{{\ttfamily arXiv:1508.02939
  [hep-th]}}.

\bibitem{Schillo:2015ssx}
M.~Schillo, E.~van~der Woerd, and T.~Wrase, ``{The general de Sitter
  supergravity component action},''
  \href{http://dx.doi.org/10.1002/prop201500074}{{\em Fortsch. Phys.}
  {\bfseries 64} (2016) 292--302},
  \href{http://arxiv.org/abs/1511.01542}{{\ttfamily arXiv:1511.01542
  [hep-th]}}.

\bibitem{Bandos:2016xyu}
I.~Bandos, M.~Heller, S.~M. Kuzenko, L.~Martucci, and D.~Sorokin, ``{The
  Goldstino brane, the constrained superfields and matter in $ \mathcal{N}=1 $
  supergravity},'' \href{http://dx.doi.org/10.1007/JHEP11(2016)109}{{\em JHEP}
  {\bfseries 11} (2016) 109}, \href{http://arxiv.org/abs/1608.05908}{{\ttfamily
  arXiv:1608.05908 [hep-th]}}.

\bibitem{Farakos:2016hly}
F.~Farakos, A.~Kehagias, D.~Racco, and A.~Riotto, ``{Scanning of the
  Supersymmetry Breaking Scale and the Gravitino Mass in Supergravity},''
  \href{http://dx.doi.org/10.1007/JHEP06(2016)120}{{\em JHEP} {\bfseries 06}
  (2016) 120}, \href{http://arxiv.org/abs/1605.07631}{{\ttfamily
  arXiv:1605.07631 [hep-th]}}.

\bibitem{Cribiori:2016qif}
N.~Cribiori, G.~Dall'Agata, F.~Farakos, and M.~Porrati, ``{Minimal Constrained
  Supergravity},'' \href{http://dx.doi.org/10.1016/j.physletb.2016.11.040}{{\em
  Phys. Lett. B} {\bfseries 764} (2017) 228--232},
  \href{http://arxiv.org/abs/1611.01490}{{\ttfamily arXiv:1611.01490
  [hep-th]}}.

\bibitem{Dudas:2015eha}
E.~Dudas, S.~Ferrara, A.~Kehagias, and A.~Sagnotti, ``{Properties of Nilpotent
  Supergravity},'' \href{http://dx.doi.org/10.1007/JHEP09(2015)217}{{\em JHEP}
  {\bfseries 09} (2015) 217}, \href{http://arxiv.org/abs/1507.07842}{{\ttfamily
  arXiv:1507.07842 [hep-th]}}.

\bibitem{Antoniadis:2015ala}
I.~Antoniadis and C.~Markou, ``{The coupling of Non-linear Supersymmetry to
  Supergravity},'' \href{http://dx.doi.org/10.1140/epjc/s10052-015-3783-0}{{\em
  Eur. Phys. J. C} {\bfseries 75} no.~12, (2015) 582},
  \href{http://arxiv.org/abs/1508.06767}{{\ttfamily arXiv:1508.06767
  [hep-th]}}.

\bibitem{DallAgata:2015pdd}
G.~Dall'Agata, S.~Ferrara, and F.~Zwirner, ``{Minimal scalar-less
  matter-coupled supergravity},''
  \href{http://dx.doi.org/10.1016/j.physletb.2015.11.066}{{\em Phys. Lett. B}
  {\bfseries 752} (2016) 263--266},
  \href{http://arxiv.org/abs/1509.06345}{{\ttfamily arXiv:1509.06345
  [hep-th]}}.

\bibitem{Ferrara:2014kva}
S.~Ferrara, R.~Kallosh, and A.~Linde, ``{Cosmology with Nilpotent
  Superfields},'' \href{http://dx.doi.org/10.1007/JHEP10(2014)143}{{\em JHEP}
  {\bfseries 10} (2014) 143}, \href{http://arxiv.org/abs/1408.4096}{{\ttfamily
  arXiv:1408.4096 [hep-th]}}.

\bibitem{Kallosh:2014via}
R.~Kallosh and A.~Linde, ``{Inflation and Uplifting with Nilpotent
  Superfields},'' \href{http://dx.doi.org/10.1088/1475-7516/2015/01/025}{{\em
  JCAP} {\bfseries 01} (2015) 025},
  \href{http://arxiv.org/abs/1408.5950}{{\ttfamily arXiv:1408.5950 [hep-th]}}.

\bibitem{Kallosh:2014hxa}
R.~Kallosh, A.~Linde, and M.~Scalisi, ``{Inflation, de Sitter Landscape and
  Super-Higgs effect},'' \href{http://dx.doi.org/10.1007/JHEP03(2015)111}{{\em
  JHEP} {\bfseries 03} (2015) 111},
  \href{http://arxiv.org/abs/1411.5671}{{\ttfamily arXiv:1411.5671 [hep-th]}}.

\bibitem{DallAgata:2014qsj}
G.~Dall'Agata and F.~Zwirner, ``{On sgoldstino-less supergravity models of
  inflation},'' \href{http://dx.doi.org/10.1007/JHEP12(2014)172}{{\em JHEP}
  {\bfseries 12} (2014) 172}, \href{http://arxiv.org/abs/1411.2605}{{\ttfamily
  arXiv:1411.2605 [hep-th]}}.

\bibitem{Carrasco:2015iij}
J.~J.~M. Carrasco, R.~Kallosh, and A.~Linde, ``{Minimal supergravity
  inflation},'' \href{http://dx.doi.org/10.1103/PhysRevD.93.061301}{{\em Phys.
  Rev. D} {\bfseries 93} no.~6, (2016) 061301},
  \href{http://arxiv.org/abs/1512.00546}{{\ttfamily arXiv:1512.00546
  [hep-th]}}.

\bibitem{DallAgata:2015zxp}
G.~Dall'Agata and F.~Farakos, ``{Constrained superfields in Supergravity},''
  \href{http://dx.doi.org/10.1007/JHEP02(2016)101}{{\em JHEP} {\bfseries 02}
  (2016) 101}, \href{http://arxiv.org/abs/1512.02158}{{\ttfamily
  arXiv:1512.02158 [hep-th]}}.

\bibitem{Dudas:2016eej}
E.~Dudas, L.~Heurtier, C.~Wieck, and M.~W. Winkler, ``{UV Corrections in
  Sgoldstino-less Inflation},''
  \href{http://dx.doi.org/10.1016/j.physletb.2016.05.072}{{\em Phys. Lett. B}
  {\bfseries 759} (2016) 121--125},
  \href{http://arxiv.org/abs/1601.03397}{{\ttfamily arXiv:1601.03397
  [hep-th]}}.

\bibitem{Delacretaz:2016nhw}
L.~V. Delacretaz, V.~Gorbenko, and L.~Senatore, ``{The Supersymmetric Effective
  Field Theory of Inflation},''
  \href{http://dx.doi.org/10.1007/JHEP03(2017)063}{{\em JHEP} {\bfseries 03}
  (2017) 063}, \href{http://arxiv.org/abs/1610.04227}{{\ttfamily
  arXiv:1610.04227 [hep-th]}}.

\bibitem{Argurio:2017joe}
R.~Argurio, D.~Coone, L.~Heurtier, and A.~Mariotti, ``{Sgoldstino-less
  inflation and low energy SUSY breaking},''
  \href{http://dx.doi.org/10.1088/1475-7516/2017/07/047}{{\em JCAP} {\bfseries
  07} (2017) 047}, \href{http://arxiv.org/abs/1705.06788}{{\ttfamily
  arXiv:1705.06788 [hep-th]}}.

\bibitem{Aldabergenov:2021obf}
Y.~Aldabergenov, A.~Chatrabhuti, and H.~Isono, ``{Nilpotent superfields for
  broken abelian symmetries},''
  \href{http://dx.doi.org/10.1140/epjc/s10052-021-09320-4}{{\em Eur. Phys. J.
  C} {\bfseries 81} no.~6, (2021) 523},
  \href{http://arxiv.org/abs/2103.11217}{{\ttfamily arXiv:2103.11217
  [hep-th]}}.

\bibitem{Aldabergenov:2021rxz}
Y.~Aldabergenov, I.~Antoniadis, A.~Chatrabhuti, and H.~Isono, ``{Quintic
  constraints for ${{\mathcal {N}}}=2$ multiplets and complete SUSY
  breaking},'' \href{http://dx.doi.org/10.1140/epjc/s10052-021-09943-7}{{\em
  Eur. Phys. J. C} {\bfseries 82} no.~1, (2022) 84},
  \href{http://arxiv.org/abs/2111.02205}{{\ttfamily arXiv:2111.02205
  [hep-th]}}.

\bibitem{Terada:2021rtp}
T.~Terada, ``{Minimal supergravity inflation without slow gravitino},''
  \href{http://dx.doi.org/10.1103/PhysRevD.103.125022}{{\em Phys. Rev. D}
  {\bfseries 103} no.~12, (2021) 125022},
  \href{http://arxiv.org/abs/2104.05731}{{\ttfamily arXiv:2104.05731
  [hep-th]}}.

\bibitem{Hasegawa:2017hgd}
F.~Hasegawa, K.~Mukaida, K.~Nakayama, T.~Terada, and Y.~Yamada, ``{Gravitino
  Problem in Minimal Supergravity Inflation},''
  \href{http://dx.doi.org/10.1016/j.physletb.2017.02.030}{{\em Phys. Lett. B}
  {\bfseries 767} (2017) 392--397},
  \href{http://arxiv.org/abs/1701.03106}{{\ttfamily arXiv:1701.03106
  [hep-ph]}}.

\bibitem{Aoki:2021nna}
S.~Aoki and T.~Terada, ``{Constrained superfields in dynamical background},''
  \href{http://dx.doi.org/10.1007/JHEP02(2022)177}{{\em JHEP} {\bfseries 02}
  (2022) 177}, \href{http://arxiv.org/abs/2111.04511}{{\ttfamily
  arXiv:2111.04511 [hep-th]}}.

\end{thebibliography}
\end{document}